\documentclass[pdflatex,sn-apa]{sn-jnl}%
\renewcommand{\pacsbullet}{\hbox{,\hskip2.5pt}}

\usepackage{graphicx}%
\usepackage{adjustbox}
\usepackage{multirow}%
\usepackage{amsmath,amssymb,amsfonts}%
\usepackage{amsthm}%
\usepackage[title]{appendix}%
\usepackage{xcolor}%

\usepackage{textcomp}%
\usepackage{manyfoot}%
\usepackage{booktabs}%
\usepackage{algorithm}%
\usepackage{algorithmicx}%
\usepackage{algpseudocode}%
\usepackage{listings}%

\usepackage{array}
\newcolumntype{L}[1]{>{\raggedright\arraybackslash}p{#1}}
\usepackage{tabularx}

\theoremstyle{thmstyleone}%
\newtheorem{theorem}{Theorem}%
\newtheorem{proposition}[theorem]{Proposition}%
\newtheorem{lemma}{Lemma}%
\newtheorem{corollary}{Corollary}%

\theoremstyle{thmstyletwo}%

\theoremstyle{thmstylethree}%

\begin{document}

\title{Computational Structure of Pairwise Housing Connectedness Networks}

\author[1]{\fnm{Seo In} \sur{Jang}}
\equalcont{These authors contributed equally to this work.}

\author[2]{\fnm{Seung Ho} \sur{Choi}}
\equalcont{These authors contributed equally to this work.}

\author[2]{\fnm{Ji-Eun} \sur{Shin}}
\equalcont{These authors contributed equally to this work.}

\author*[3]{\fnm{Hyunwoo} \sur{Lee}}
\email{hyunwoolee@kias.re.kr}

\affil[1]{\orgdiv{Department of Statistics},
\orgname{Kyungpook National University},
\orgaddress{\street{80, Daehak-ro, Buk-gu},
\postcode{41566},
\city{Daegu},
\country{Republic of Korea}}}

\affil[2]{\orgdiv{Department of Mathematics},
\orgname{Kyungpook National University},
\orgaddress{\street{80, Daehak-ro, Buk-gu},
\postcode{41566},
\city{Daegu},
\country{Republic of Korea}}}

\affil*[3]{\orgname{Korea Institute for Advanced Study},
\orgaddress{\street{85 Hoegiro, Dongdaemun-gu},
\postcode{02455},
\city{Seoul},
\country{Republic of Korea}}}

\abstract
{Directed housing-submarket networks combine common market movement with heterogeneous bilateral relations, and different node-level measures need not yield distinct empirical rankings. We construct a positive pairwise net-connectedness network from the generalized forecast error variance decomposition (GFEVD) of a vector autoregression (VAR) and use it to compare complementary node positions. Net directional connectedness (NET), PageRank, and Viral Centrality summarize net directional balance, recursively weighted incoming position, and normalized multistep reach. When bilateral net differences are nonzero, the network is a weighted tournament. A least-squares edge-flow decomposition separates the pairwise-net matrix into a NET-equivalent gradient component and a NET-orthogonal residual, while a monotone-allocation condition establishes exact PageRank--indegree correspondence in a transitive tournament. In Seoul's 25-district network, the gradient component accounts for 70.9\% of the squared Frobenius norm, leaving 29.1\% in the residual component. PageRank is strongly aligned with indegree (Spearman $\rho=0.987$), and Viral Centrality with outdegree ($\rho=0.935$); the PageRank alignment exceeds the range of independently weighted simulations with comparable tournament consistency and edge-weight variation. Three common-component treatments reduce total connectedness from 81.98\% to 50.60--52.00\%, a 36.6--38.3\% reduction relative to the raw system, and reverse 97--98 of 300 pairwise net directions. Thus, both the magnitude and the apparent direction of connectedness depend materially on the treatment of citywide common movement. Songpa's positive NET position remains prominent across common-component treatments and dynamic specifications and ranks first in both chronological subsamples.}



\maketitle

\section{Introduction}
\label{sec:introduction}

Housing prices across local submarkets move jointly as demand, credit conditions, and expectations affect several markets at once~\citep{meen1999regional,favara2015credit,piazzesi2020segmented,bailey2018economic,burnside2016understanding}. Understanding this dependence requires examining both aggregate connectedness and the positions of individual districts. The generalized forecast error variance decomposition (GFEVD) of a vector autoregression (VAR) provides a reduced-form representation of directional connectedness~\citep{diebold2009measuring,diebold2012better,diebold2014network}. Applications span regional, cross-city, and intraurban housing markets~\citep{tsai2015spillover,zhang2020detecting,chen2023measuring,gabauer2024estimating,kim2021micro,hwang2021analyzing,lee2019cross}. Here, spillover and transmission refer to forecast-error variance attribution rather than structurally identified causal effects~\citep{lanne2016generalized}.

Two interpretation problems arise before network rankings are assigned economic roles. First, connectedness estimated from the raw district panel may reflect citywide movement shared across submarkets as well as bilateral relations among them. Second, even after bilateral netting, the pairwise directional structure need not be reducible to a single upstream--downstream ordering. Net directional connectedness (NET) summarizes the balance between connectedness transmitted and received, while PageRank incorporates recursively weighted incoming links and Viral Centrality (VC) summarizes normalized reach along successive directed paths. Although these measures differ mathematically, their rankings can closely reproduce simpler measures such as the number or total weight of incoming and outgoing links. The interpretation of a prominent district therefore depends both on what part of the pairwise network is represented by the chosen measure and on how the estimated position changes with common movement, model dynamics, or the sample period.

We develop a computational framework for making these assessments on the positive pairwise net-connectedness network. When every bilateral net difference is nonzero, the network is a weighted tournament: each pair of districts is connected in exactly one direction. A least-squares edge-flow decomposition characterizes what the scalar NET representation retains from the pairwise-net matrix. On the complete district graph, its gradient component is in one-to-one correspondence with NET and represents the part of the pairwise flow that can be summarized by a single potential ordering; the orthogonal residual contains directional variation that NET does not represent. A monotone-allocation condition then provides a sufficient case for exact PageRank--indegree rank correspondence in a transitive tournament. We examine departures from the transitive uniform-weight special case in 50,000 controlled weighted-tournament simulations. Varying edge directions and weight dispersion shows how these features affect degree--centrality alignment. Comparisons with degree, strength, and uniform-weight rankings identify where the empirical ordering follows simpler connectivity measures and where individual districts depart from them.

The empirical workflow combines this structural assessment with three further diagnostics: alternative common-component treatments, alternative VAR specifications, and a chronological sample split. It records changes across all districts and follows prominent candidates through the applicable assessments. The temporal analysis reports observed rank persistence separately from bootstrap reproduction frequencies. Together, the diagnostics record whether prominence is reproduced by simpler connectivity and how it changes with common movement, dynamic specification, or sample timing.

The application uses weekly apartment-price indices for Seoul's 25 districts from May 2012 to August 2026. The NET-equivalent gradient component accounts for 70.9\% of the squared Frobenius norm of the pairwise-net matrix, while 29.1\% remains in the NET-orthogonal residual. PageRank and VC closely follow indegree and outdegree, with rank correlations of 0.987 and 0.935, respectively, and Seoul's PageRank alignment exceeds that of the simulated networks matched on tournament consistency and weight dispersion under independent weight assignment. Common-component treatments reduce total connectedness from 81.98\% to 50.60--52.00\%, equivalent to a 36.6--38.3\% reduction from the raw system, and reverse 97--98 of 300 pairwise directions. These changes show that the raw network should not be read mechanically as a fixed map of district-to-district ripple effects. Songpa nevertheless retains a prominent positive NET position across the examined treatments and VAR specifications and ranks first in both observed chronological subsamples; its bootstrap reproducibility is weaker in the early subsample.

Section~\ref{sec:literature} reviews the related literature. Section~\ref{sec:methods} presents the network construction and diagnostic workflow, and Section~\ref{sec:results} reports the results. Sections~\ref{sec:discussion} and~\ref{sec:conclusion} discuss their implications and conclude.  

\section{Related Literature and Analytical Positioning}
\label{sec:literature}

\subsection{Housing spillovers and intraurban connectedness}
\label{subsec:housing_spillovers}

Regional housing markets are economically interdependent, and local price movements can coexist with broader changes in credit conditions, expectations, and housing demand. These interactions have long been studied through ripple effects, spatial dependence, and directional spillovers. \citet{meen1999regional} provides a foundational account of regional house-price diffusion, while subsequent studies document spillovers across regional, cross-city, and intraurban housing markets~\citep{tsai2015spillover,zhang2020detecting,chen2023measuring,gabauer2024estimating}. The ripple-effect perspective also raises a structural question: whether estimated bilateral directions can be organized by a coherent upstream--downstream ordering, or whether part of the directional pattern remains outside such an ordering. Forecast-error variance decompositions provide a natural framework for representing these interactions as a directed network and for comparing the relative positions of individual markets within a common system~\citep{diebold2009measuring,diebold2012better,diebold2014network}.

Evidence from Korea likewise documents substantial cross-regional dependence and time-varying directional connectedness~\citep{lee2019cross,alyahyaee2021korean,park2024study}. At a finer spatial scale, \citet{kim2021micro} report spillovers among selected submarkets in the Seoul metropolitan area, while \citet{hwang2021analyzing} use a time-varying parameter VAR to trace connectedness across Seoul's 25 districts. \citet{lee2019cross} document changes in connectedness patterns across Korean regional housing markets, and \citet{payne2023timevarying} report substantial time variation across U.S. metropolitan housing markets. Regional housing prices are also jointly exposed to common conditions, with \citet{holly2010spatio} showing that common factors and spatial dependence can both matter for regional house-price dynamics. This creates a second interpretation issue for connectedness networks: a raw network can combine bilateral directional dependence with movement shared across the wider market. The present study therefore examines both the structural content of pairwise net connectedness and the sensitivity of its magnitude, directions, and node positions to alternative treatments of common movement, dynamic specification, and sample timing.

Directional connectedness contains more information than can be represented by a single node-level NET value. NET provides an economically useful summary of the balance between connectedness transmitted to and received from the rest of the system, but this compression does not fully describe a district's position within the directed topology. Districts with similar NET values can differ in their incoming structure, outgoing structure, and positions along multistep directed paths. The full connectedness network therefore permits a richer characterization of district-level positions than the net balance alone. At the same time, a high ranking in one estimated network need not have the same interpretation under alternative empirical conditions. This motivates extending scalar NET with recursive and multistep positions while treating their structural distinctiveness and condition sensitivity as quantities to be assessed rather than assumed.

\subsection{Network measures and computational interpretation}
\label{subsec:recursive_centrality}

In the connectedness literature, NET is conventionally used to summarize the balance between connectedness transmitted to and received from the rest of the system, with positive and negative values interpreted as net-transmitter and net-receiver positions \citep{diebold2012better,diebold2014network,lee2019cross}. PageRank and related centrality measures have also been applied to variance-decomposition and systemic-risk networks, where recursive or path-based rankings can remain close to simpler centrality or correlation-based orderings \citep{chanlau2018systemic,dungey2018systemic,vandeleur2017network}. Unlike these empirical comparisons, the present pairwise-netted network has a tournament structure that provides a structural basis for such alignment, while allowing node-level departures and condition sensitivity to be quantified. PageRank and Viral Centrality are used here on the pairwise-netted topology without assigning additional economic role labels a priori. They summarize recursive incoming structure and multistep directed-path information that are not determined by the scalar NET balance alone.

Different mathematical measures, however, need not provide distinct empirical rankings. When no pairwise net difference is exactly zero, the support of the positive part of a skew-symmetric pairwise net matrix defines a tournament, while the positive entries retain the corresponding edge weights. Tournament structure can support strong alignment between recursive or multistep centrality measures and simpler degree or strength rankings, but the degree of alignment need not be determined by tournament consistency alone because edge-weight heterogeneity and the detailed arrangement of directions and weights also matter~\citep{schoch2017correlations,perra2008spectral}. The analysis therefore quantifies how closely the PageRank and Viral Centrality rankings reproduce these simpler structural hierarchies and where node-level departures remain.

Factor-based and bank-level network studies show that common conditions can materially affect inferred network structure~\citep{barigozzi2017network,hale2019monitoring}, while regularized VAR methods have been developed for high-dimensional connectedness estimation~\citep{demirer2018estimating,gabauer2024estimating}. These strands of literature motivate a broader diagnostic layer for assessing how the interpretation of district-level network positions depends on common movement, dynamic specification, and sample timing, and how recursive or multistep positions align with simpler structural hierarchies.

Recent work in \textit{Computational Economics} combines Diebold--Yilmaz spillover measures with complex-network analysis to characterize directional transmission patterns and transmitter--receiver roles across interconnected markets~\citep{fu2026dynamic}. Related work uses dynamic Bayesian networks together with local and global network measures to characterize time-varying financial interdependence and network structure~\citep{bernardelli2026dynamic}. Higher-order approaches further extend pairwise network analysis by combining directional information-flow networks, PageRank-based risk centrality, and hypergraph representations of multi-entity contagion \citep{akguller2026detecting}.

Building on these strands, the present study extends pairwise VAR--GFEVD net-connectedness analysis beyond scalar NET by computing PageRank and Viral Centrality on the positive pairwise net-connectedness topology obtained after bilateral netting. To characterize the information represented by scalar NET, we specialize the least-squares edge-flow decomposition of \citet{jiang2011statistical} to the complete pairwise-net matrix. The decomposition links NET to a global gradient ordering and separates the residual bilateral structure not represented by that ordering. A transitive-tournament benchmark and controlled weighted-tournament simulations then provide structural reference points for interpreting degree--centrality alignment, while the broader diagnostic layer evaluates how prominent positions change with common movement, dynamic specification, and sample timing.

\section{Data and Empirical Framework}
\label{sec:methods}

\subsection{Seoul housing data}
\label{subsec:seoul_data}

The main empirical analysis uses weekly apartment sale-price indices for Seoul's 25 administrative districts (\textit{gu}) from the Weekly Apartment Price Trends component of the National Housing Price Trend Survey compiled by the Korea Real Estate Board (REB), obtained through the R-ONE Real Estate Statistics Information System.\footnote{\url{https://www.reb.or.kr/r-one/portal/stat/easyStatPage/T244183132827305.do} (accessed August 31, 2026). The REB constructs the apartment price indices using the Jevons index methodology and publishes the weekly series at the city and district levels.} The REB notes that its housing-price statistics have undergone periodic sample adjustments, expansions, and redesigns. These changes are not separately adjusted in the present analysis, and district-level sample sizes and index precision are not publicly reported.

For each district, let $p_{i,t}$ denote the index level of district $i$ in week $t$. The weekly housing return is defined as

\begin{equation}
r_{i,t}
=
\frac{p_{i,t}-p_{i,t-1}}{p_{i,t-1}},
\label{eq:housing_return}
\end{equation}

where $r_{i,t}$ is expressed in decimal units. All VAR--GFEVD estimations use these decimal returns. For descriptive presentation only, Table~\ref{tab1} reports return statistics on the percentage scale, $100r_{i,t}$.

The source index panel begins on May 7, 2012, and ends on August 24, 2026. Because Eq.~\eqref{eq:housing_return} requires a lagged index value, the first return observation is dated May 14, 2012. Three weekly dates, October 2, 2017, January 27, 2025, and October 6, 2025, are absent from the source panel. No interpolation or other imputation is applied, so the returns immediately following these gaps span two calendar weeks and are retained as such. The resulting panel contains 743 return observations per district. A sensitivity check excluding the three 14-day returns and preventing VAR lags from crossing the corresponding gaps is reported in Supplementary Section~S1.1.

After accounting for the single lag, the baseline VAR(1) estimation uses 742 effective observations. Table~\ref{tab1} reports descriptive statistics for the district-level return series.

\begin{table}[!htbp]
\centering
\caption{Summary statistics and preliminary diagnostics for weekly district-level housing returns}
\label{tab1}

\renewcommand{\arraystretch}{0.95}

\vspace{2pt}
\noindent\textit{Panel A. Summary statistics}

\vspace{1pt}

\begin{tabular*}{\linewidth}{@{\extracolsep{\fill}}lccccc@{}}
\toprule
Variable & Obs. & Mean & Std. Dev. & Skewness & Kurtosis \\
\midrule
Gangnam (GN) & 743 & 0.064 & 0.212 & -0.32 & 8.16 \\
Gangdong (GD) & 743 & 0.059 & 0.215 & 0.13 & 7.79 \\
Gangbuk (GB) & 743 & 0.019 & 0.148 & -2.14 & 15.52 \\
Gangseo (GS) & 743 & 0.043 & 0.166 & -1.26 & 7.02 \\
Gwanak (GA) & 743 & 0.038 & 0.147 & -1.35 & 11.74 \\
Gwangjin (GJ) & 743 & 0.052 & 0.167 & 1.14 & 15.92 \\
Guro (GU) & 743 & 0.045 & 0.137 & -0.63 & 5.69 \\
Geumcheon (GC) & 743 & 0.026 & 0.146 & -2.45 & 17.98 \\
Nowon (NW) & 743 & 0.032 & 0.178 & -2.49 & 18.11 \\
Dobong (DB) & 743 & 0.008 & 0.177 & -2.53 & 16.36 \\
Dongdaemun (DDM) & 743 & 0.042 & 0.162 & -1.42 & 9.95 \\
Dongjak (DJ) & 743 & 0.059 & 0.163 & -0.49 & 8.10 \\
Mapo (MP) & 743 & 0.066 & 0.167 & 0.26 & 14.31 \\
Seodaemun (SDM) & 743 & 0.041 & 0.154 & -1.67 & 12.23 \\
Seocho (SC) & 743 & 0.068 & 0.169 & 0.73 & 5.44 \\
Seongdong (SD) & 743 & 0.074 & 0.198 & 1.46 & 11.44 \\
Seongbuk (SB) & 743 & 0.037 & 0.158 & -1.80 & 14.94 \\
Songpa (SP) & 743 & 0.078 & 0.227 & 0.56 & 8.02 \\
Yangcheon (YC) & 743 & 0.054 & 0.178 & 0.43 & 8.56 \\
Yeongdeungpo (YDP) & 743 & 0.058 & 0.163 & -0.83 & 7.19 \\
Yongsan (YS) & 743 & 0.042 & 0.164 & 0.33 & 9.71 \\
Eunpyeong (EP) & 743 & 0.026 & 0.139 & -2.51 & 16.81 \\
Jongno (JR) & 743 & 0.036 & 0.115 & -1.36 & 11.15 \\
Jung (JG) & 743 & 0.036 & 0.181 & -1.97 & 13.90 \\
Jungnang (JN) & 743 & 0.023 & 0.144 & -1.85 & 15.93 \\
\bottomrule
\end{tabular*}

\vspace{4pt}

\noindent\textit{Panel B. Stationarity, distributional, and lag-order diagnostics}

\vspace{1pt}

\begin{tabularx}{\linewidth}{@{}p{0.15\linewidth} p{0.29\linewidth} X@{}}
\toprule
Diagnostic & Null hypothesis & Result \\
\midrule
ADF  & Unit root & 25 of 25 rejected at 1\% \\
PP   & Unit root & 25 of 25 rejected at 1\% \\
KPSS & Level stationarity & 2 of 25 rejected at 5\%, 0 after BH adjustment \\
JB   & Normal distribution & 25 of 25 rejected at 5\%, 25 after BH adjustment \\
BIC  & Lag-order selection & $p=1$, BIC $=-362.95$ \\
\bottomrule
\end{tabularx}

\vspace{2pt}

\begin{minipage}{\linewidth}
\scriptsize
\raggedright
\textit{Note}. Return statistics are reported on the percentage scale, $100r_{i,t}$, while VAR--GFEVD estimation uses decimal simple returns $r_{i,t}$. Kurtosis is raw kurtosis. ADF and PP test a unit-root null, KPSS tests level stationarity, and JB tests normality. BH denotes Benjamini--Hochberg adjustment across 25 district-level tests. Detailed test statistics and lag-order results are reported in Supplementary Section~S1.
\end{minipage}

\end{table}

The return observations are collected in the matrix

\begin{equation}
\mathbf{R}
=
\begin{bmatrix}
r_{1,1} & r_{2,1} & \cdots & r_{N,1} \\
r_{1,2} & r_{2,2} & \cdots & r_{N,2} \\
\vdots & \vdots & \ddots & \vdots \\
r_{1,T} & r_{2,T} & \cdots & r_{N,T}
\end{bmatrix}
\in \mathbb{R}^{T\times N},
\qquad
T=743,\quad N=25.
\label{eq:return_matrix}
\end{equation}

This matrix provides the common empirical input for the connectedness and network calculations below.

\subsection{Stationarity tests and lag-order selection}
\label{subsec:stationarity}

Before estimating the VAR, we examine the weekly return series using Augmented Dickey--Fuller (ADF)~\citep{dickey1979distribution}, Phillips--Perron (PP)~\citep{phillips1988testing}, and KPSS tests~\citep{kwiatkowski1992testing}. The baseline ADF and PP tests are specified without a deterministic term, consistent with the return-based VAR specification. The PP test uses the $Z_{\tau}$ statistic with a Newey--West long-run variance correction~\citep{newey1987simple}, with bandwidth selected as $\lceil 12(T/100)^{1/4}\rceil$, yielding 20 for Seoul. Both tests reject the unit-root null for all 25 district series at the 1\% level. The KPSS test, which takes level stationarity as the null, rejects that null for 2 of the 25 series at the 5\% level using unadjusted $p$-values and for none after Benjamini--Hochberg adjustment~\citep{benjamini1995controlling}. The Jarque--Bera test~\citep{jarque1987test} rejects normality for all 25 series under both unadjusted and adjusted $p$-values. These results indicate broadly stationary but non-normal weekly returns. Detailed results are reported in Table~\ref{tab1} and Supplementary Section~S1.

The VAR lag order is selected by the Bayesian information criterion (BIC)~\citep{schwarz1978estimating} over $p=1,\ldots,10$, with the minimum at $p=1$, which is retained as the reference specification. Because residual dependence remains, the resulting VAR--GFEVD networks are interpreted as model-conditional reduced-form summaries. VAR(4), which shows substantially lower equation-level residual autocorrelation than VAR(1), is used as a lower-residual-autocorrelation benchmark in the full-sample specification analysis rather than as a replacement reference model; the remaining joint dependence prevents its interpretation as a fully whitened or uniquely preferred dynamic representation.

\subsection{VAR--GFEVD connectedness}
\label{subsec:gfevd}

Directional connectedness is estimated using generalized forecast-error variance decomposition (GFEVD) within a vector autoregressive model, following \citet{diebold2012better,diebold2014network}. Let $\mathbf{r}_t=(r_{1,t},r_{2,t},\ldots,r_{N,t})^\top$ denote the $N$-dimensional return vector. We consider the covariance-stationary VAR($p$)

\begin{equation}
\mathbf{r}_t
=
\sum_{\ell=1}^{p}
\mathbf{\Phi}_{\ell}\mathbf{r}_{t-\ell}
+
\boldsymbol{\varepsilon}_t,
\label{eq:var_model}
\end{equation}

where $\mathbf{\Phi}_{\ell}$ is the $N\times N$ coefficient matrix and $\boldsymbol{\varepsilon}_t$ has variance--covariance matrix $\mathbf{\Sigma}$. The reference model uses $p=1$ and is estimated without an intercept or other deterministic terms. Because deterministic-term treatment can affect the estimated network, an intercept-included VAR(1) is also evaluated as a sensitivity specification and reported in the Supplementary Material.

The VAR representation can be written in moving-average form as

\begin{equation}
\mathbf{r}_t
=
\sum_{h=0}^{\infty}
\mathbf{A}_h
\boldsymbol{\varepsilon}_{t-h},
\label{eq:ma_representation}
\end{equation}

where $\mathbf{A}_0=\mathbf{I}_N$ and the subsequent $\mathbf{A}_h$ matrices are obtained recursively from the VAR coefficients.

Following \citet{koop1996impulse} and \citet{pesaran1998generalized}, the $H$-step-ahead generalized forecast-error variance contribution of shocks in submarket $j$ to the forecast-error variance of submarket $i$ is

\begin{equation}
\theta^{g}_{ij}(H)
=
\frac{
\sigma_{jj}^{-1}
\sum_{h=0}^{H-1}
\left(
\mathbf{e}_{i}^{\top}
\mathbf{A}_h
\mathbf{\Sigma}
\mathbf{e}_{j}
\right)^2
}{
\sum_{h=0}^{H-1}
\left(
\mathbf{e}_{i}^{\top}
\mathbf{A}_h
\mathbf{\Sigma}
\mathbf{A}_h^{\top}
\mathbf{e}_{i}
\right)
},
\label{eq:gfevd}
\end{equation}

where $\sigma_{jj}$ is the $j$th diagonal element of $\mathbf{\Sigma}$ and $\mathbf{e}_i$ is the corresponding selection vector. Because generalized variance-decomposition entries do not necessarily sum to one across rows, they are normalized as

\begin{equation}
\widetilde{\theta}^{g}_{ij}(H)
=
\frac{
\theta^{g}_{ij}(H)
}{
\sum_{k=1}^{N}
\theta^{g}_{ik}(H)
},
\qquad
\sum_{j=1}^{N}
\widetilde{\theta}^{g}_{ij}(H)
=
1.
\label{eq:normalized_gfevd}
\end{equation}

The normalized GFEVD matrix is

\begin{equation}
\widetilde{\mathbf{\Theta}}^{g}(H)
=
\left[
\widetilde{\theta}^{g}_{ij}(H)
\right]_{i,j=1}^{N},
\label{eq:gfevd_matrix}
\end{equation}

from which directional connectedness is obtained. The connectedness received by submarket $i$ from all other submarkets is

\begin{equation}
FROM_i(H)
=
100
\sum_{\substack{j=1 \\ j\neq i}}^{N}
\widetilde{\theta}^{g}_{ij}(H),
\label{eq:from}
\end{equation}

whereas connectedness transmitted by submarket $i$ to the remaining system is

\begin{equation}
TO_i(H)
=
100
\sum_{\substack{j=1 \\ j\neq i}}^{N}
\widetilde{\theta}^{g}_{ji}(H).
\label{eq:to}
\end{equation}

The corresponding net directional connectedness is

\begin{equation}
NET_i(H)
=
TO_i(H)-FROM_i(H).
\label{eq:net}
\end{equation}

A positive $NET_i(H)$ indicates that submarket $i$ transmits more forecast-error variance to the rest of the system than it receives, whereas a negative value indicates the reverse balance. These positions are conventionally described as net-transmitter and net-receiver positions, respectively.

System-wide connectedness is summarized by

\begin{equation}
TCI(H)
=
\frac{100}{N}
\sum_{\substack{i,j=1 \\ i\neq j}}^{N}
\widetilde{\theta}^{g}_{ij}(H).
\label{eq:tci}
\end{equation}

The baseline analysis uses $H=10$ as the reference horizon. With weekly data, this represents forecast-error variance accumulated over ten forecast steps and should not be interpreted as a calendar-time lead. Alternative horizons are examined in the supplementary sensitivity analysis.

\subsection{Directed pairwise net-connectedness, NET, and edge-flow decomposition}
\label{subsec:net_network}

To represent bilateral net direction explicitly, the pairwise net connectedness from $i$ to $j$ is defined as

\begin{equation}
N_{ij}(H)
=
\widetilde{\theta}^{g}_{ji}(H)
-
\widetilde{\theta}^{g}_{ij}(H).
\label{eq:pairwise_net}
\end{equation}

Thus, $N_{ij}(H)>0$ indicates that the estimated net direction is from $i$ to $j$. By construction,

\begin{equation}
N_{ji}(H)=-N_{ij}(H),
\label{eq:net_skew}
\end{equation}

so $\mathbf{N}(H)=[N_{ij}(H)]$ is skew-symmetric. We retain the positive direction of each pair through

\begin{equation}
A^{+}_{ij}(H)
=
\max\left\{N_{ij}(H),0\right\},
\label{eq:positive_net}
\end{equation}

and define

\begin{equation}
\mathbf{A}^{+}(H)
=
\left[A^{+}_{ij}(H)\right]_{i,j=1}^{N}.
\label{eq:adjacency_positive}
\end{equation}

An edge $i\rightarrow j$ therefore represents the estimated net direction for that pair, with weight equal to the corresponding net difference. The resulting $\mathbf{A}^{+}(H)$ is the positive pairwise net-connectedness network rather than the original gross GFEVD network and provides the common directed topology for the node-position measures used below. In the baseline Seoul network, no bilateral net difference is exactly zero, so the positive network forms a complete weighted tournament. For this tournament, cyclic organization is summarized by the Kendall--Babington Smith consistency coefficient $\eta_{\mathrm{KBS}}=1-c_3/c_{3,\max}$, where $c_3$ is the number of cyclic triads and $c_{3,\max}=(N^3-N)/24$ for odd $N$ \citep{kendall1940paired}. The coefficient is used as a global summary of tournament consistency rather than as a sufficient determinant of centrality--degree alignment.

The network construction is directly related to NET:

\begin{equation}
NET_i(H)
=
100
\sum_{j=1}^{N}
N_{ij}(H),
\label{eq:net_pairwise_identity}
\end{equation}

and, because $N_{ij}(H)=A^{+}_{ij}(H)-A^{+}_{ji}(H)$,

\begin{equation}
NET_i(H)
=
100
\left[
\sum_{j=1}^{N}A^{+}_{ij}(H)
-
\sum_{j=1}^{N}A^{+}_{ji}(H)
\right].
\label{eq:net_strength_identity}
\end{equation}

Define outgoing and incoming net-connectedness strengths as

\begin{equation}
s_i^{\mathrm{out}}(H)
=
\sum_{j=1}^{N}A^{+}_{ij}(H),
\qquad
s_i^{\mathrm{in}}(H)
=
\sum_{j=1}^{N}A^{+}_{ji}(H).
\label{eq:network_strengths}
\end{equation}

Then

\begin{equation}
NET_i(H)
=
100
\left[
s_i^{\mathrm{out}}(H)
-
s_i^{\mathrm{in}}(H)
\right].
\label{eq:net_strength}
\end{equation}

Specializing the least-squares edge-flow decomposition to the complete pairwise-net matrix yields the following result \citep{jiang2011statistical}.

\begin{lemma}[NET and the least-squares gradient component]
\label{lem:net_projection}
Let
\[
p_i(H)=\frac{NET_i(H)}{100N},
\qquad
G_{ij}(H)=p_i(H)-p_j(H).
\]
Among all vectors $\mathbf{x}$ satisfying $\sum_i x_i=0$, $\mathbf{p}(H)$ uniquely minimizes
\[
\sum_{i<j}
\left[
N_{ij}(H)-(x_i-x_j)
\right]^2.
\]
Writing
\[
\mathbf{N}(H)=\mathbf{G}(H)+\mathbf{C}(H),
\]
the two components are orthogonal under the Frobenius inner product, and
\[
\sum_{j=1}^{N}C_{ij}(H)=0
\qquad
\text{for every }i.
\]
Hence, $\mathbf{G}(H)$ is in one-to-one correspondence with NET, whereas $\mathbf{C}(H)$ is a divergence-free, NET-orthogonal residual component containing the pairwise-net matrix variation not represented by NET.
\end{lemma}

The relative magnitude of the NET-equivalent component is summarized by
\[
\psi_G(H)
=
\frac{\|\mathbf{G}(H)\|_F^2}
{\|\mathbf{N}(H)\|_F^2}.
\]
By orthogonality, the corresponding residual share is
\[
\psi_C(H)
=
1-\psi_G(H)
=
\frac{\|\mathbf{C}(H)\|_F^2}
{\|\mathbf{N}(H)\|_F^2}.
\]
The two shares are used as descriptive summaries of how much of the weighted pairwise-net structure is represented by the NET-equivalent gradient and how much remains orthogonal to it. Economically, the gradient component provides a compact upstream--downstream representation of the bilateral net flow, whereas the residual records directional variation that cannot be summarized by that single ordering. This interpretation is reduced-form: neither component is treated as an identified causal propagation mechanism.

For the recursive and multistep measures below, we distinguish the forward normalization from the dangling-node correction used by PageRank. For nodes with positive outgoing strength,

\begin{equation}
B_{ij}(H)
=
\frac{A^{+}_{ij}(H)}
{\sum_{k=1}^{N}A^{+}_{ik}(H)}
=
\frac{A^{+}_{ij}(H)}
{s_i^{\mathrm{out}}(H)}.
\label{eq:cascade_normalization}
\end{equation}

For a zero-outdegree node,

\begin{equation}
B_{ij}(H)=0,
\qquad j=1,\ldots,N.
\label{eq:cascade_sink}
\end{equation}

Hence, $\mathbf{B}(H)$ preserves such nodes as sinks. For PageRank, the corresponding dangling row is instead completed uniformly,

\begin{equation}
Q_{ij}(H)
=
\begin{cases}
B_{ij}(H), & s_i^{\mathrm{out}}(H)>0,\\[1mm]
\dfrac{1}{N}, & s_i^{\mathrm{out}}(H)=0.
\end{cases}
\label{eq:pagerank_transition}
\end{equation}

so that

\begin{equation}
\sum_{j=1}^{N}Q_{ij}(H)=1
\qquad
\text{for all }i.
\label{eq:q_row_stochastic}
\end{equation}

On non-dangling rows, both matrices preserve the estimated pairwise directions. $\mathbf{B}(H)$ is used for Viral Centrality, whereas $\mathbf{Q}(H)$, including its uniform dangling-row completion, provides the row-stochastic transition matrix used for PageRank.

\subsection{PageRank and recursively weighted incoming position}
\label{subsec:pagerank}

PageRank is applied to $\mathbf{Q}(H)$ to summarize recursively weighted incoming position. Let $\alpha\in(0,1)$ denote the damping parameter. Following \citet{brin1998anatomy}, define

\begin{equation}
\mathbf{G}_{\alpha}(H)
=
\alpha \mathbf{Q}(H)
+
(1-\alpha)
\frac{1}{N}
\mathbf{1}\mathbf{1}^{\top},
\label{eq:google_matrix}
\end{equation}

where $\mathbf{1}$ is the $N\times1$ vector of ones. The PageRank vector is the normalized left stationary vector of $\mathbf{G}_{\alpha}(H)$,

\begin{equation}
\mathbf{PR}(H)^{\top}
=
\mathbf{PR}(H)^{\top}
\mathbf{G}_{\alpha}(H),
\label{eq:pagerank_stationary}
\end{equation}

subject to

\begin{equation}
PR_i(H)\geq0,
\qquad
\sum_{i=1}^{N}PR_i(H)=1.
\label{eq:pagerank_normalization}
\end{equation}

Equivalently,

\begin{equation}
PR_i(H)
=
\frac{1-\alpha}{N}
+
\alpha
\sum_{j=1}^{N}
PR_j(H)Q_{ji}(H).
\label{eq:pagerank_recursive}
\end{equation}

Equation~\eqref{eq:pagerank_recursive} shows that PageRank summarizes a recursively weighted incoming position: a district receives a larger score when it receives greater transition weight from nodes that themselves have high PageRank. Terms introduced by the uniform dangling-node correction do not represent estimated spillover links. Accordingly, PageRank is interpreted as a recursively weighted incoming network position.

PageRank differs from NET in definition because it summarizes recursively weighted incoming position rather than net directional balance. A district may therefore have a negative NET and relatively high PageRank, or a positive NET and relatively low PageRank. The extent to which the PageRank ordering departs from simpler incoming hierarchies is assessed in Section~\ref{subsec:diagnostic_framework}.

The baseline damping parameter is

\begin{equation}
\alpha=0.85,
\label{eq:pagerank_alpha}
\end{equation}

with alternative damping values reported in the supplementary sensitivity analysis.

A transitive tournament with monotone outgoing allocation provides a sufficient condition for exact PageRank--indegree alignment.

\begin{proposition}[Monotone-allocation benchmark]
\label{prop:pagerank_indegree_boundary}
Consider a complete transitive tournament with strictly positive edge weights, whose nodes are indexed so that $j\rightarrow k$ whenever $j<k$. Let $\mathbf{Q}$ be formed using row normalization and the uniform dangling-node correction. Suppose that
\[
Q_{j,k+1}\geq Q_{j,k}
\qquad
\text{for every }j<k<N.
\]
Under uniform PageRank teleportation with $\alpha\in(0,1)$, PageRank is strictly increasing in indegree and therefore induces the same node ranking as indegree.
\end{proposition}

\begin{corollary}[Exact ordering under a zero residual component]
\label{cor:net_pagerank_alignment}
Suppose that $\mathbf{C}(H)=\mathbf{0}$ in Lemma~\ref{lem:net_projection} and that all NET values are distinct. Then the PageRank ranking coincides with the indegree ranking and is exactly the reverse of the NET ranking. Consequently, $\mathbf{C}(H)\neq\mathbf{0}$ is necessary for a departure from this ordering, but it is not sufficient.
\end{corollary}

\begin{corollary}[Gradient share of a unit-weight tournament]
\label{cor:unit_weight_gradient_share}
For a complete tournament on an odd number $N\geq3$ of nodes with unit positive edge weights, the gradient share satisfies
\[
\psi_G
=
\eta_{\mathrm{KBS}}
\frac{2(N+1)}{3N}.
\]
Hence, a transitive uniform-weight tournament has
\[
\psi_G
=
\frac{2(N+1)}{3N}.
\]
For $N=25$, this transitive uniform-weight value is $0.6933$.
\end{corollary}

For $N>2$, a transitive uniform-weight tournament illustrates why the converse does not hold: it can have $\mathbf{C}(H)\neq\mathbf{0}$ while satisfying the monotone-allocation condition and retaining exact PageRank--indegree rank correspondence.

Derivations of Lemma~\ref{lem:net_projection}, Proposition~\ref{prop:pagerank_indegree_boundary}, and Corollaries~\ref{cor:net_pagerank_alignment} and~\ref{cor:unit_weight_gradient_share} are provided in Supplementary Section~S5.1. To examine departures from the transitive uniform-weight special case, we simulate complete weighted tournaments with $N=25$ by independently perturbing pairwise orientations and varying the coefficient of variation of positive edge weights. The design comprises 50 parameter combinations and 1,000 replications per combination; its full specification and additional results are reported in Supplementary Section~S5.2. The resulting benchmark is compared with the empirical network in Stage~2.

\subsection{Viral Centrality and normalized multistep reach}
\label{subsec:viral_centrality}

Viral Centrality is implemented here as a deterministic recursion that summarizes normalized multistep reach from each seed through successive directed paths~\citep{fink2023centrality}. The calculation uses the forward matrix $\mathbf{B}(H)$ defined in Eqs.~\eqref{eq:cascade_normalization}--\eqref{eq:cascade_sink}, with

\begin{equation}
\beta_{ij}(H)=B_{ij}(H)
\label{eq:vc_edge_probability}
\end{equation}

as the normalized forward weight associated with edge $i\rightarrow j$. These weights are not interpreted as empirically calibrated diffusion probabilities, and the resulting Viral Centrality score is not an exact expected cascade size or an estimate of economic propagation.

For a process seeded at node $s$, let $p_i^{\mathrm{sus}}(t)$ and $p_i^{\mathrm{act}}(t)$ denote its susceptible and newly activated states at step $t$. Initialization is

\begin{equation}
p_i^{\mathrm{sus}}(0)
=
\begin{cases}
0, & i=s,\\
1, & i\neq s,
\end{cases}
\qquad
p_i^{\mathrm{act}}(0)
=
\begin{cases}
1, & i=s,\\
0, & i\neq s.
\end{cases}
\label{eq:vc_initialization}
\end{equation}

The escape factor is

\begin{equation}
p_i^{\mathrm{esc}}(t)
=
\prod_{j=1}^{N}
\left[
1-p_j^{\mathrm{act}}(t-1)\beta_{ji}(H)
\right],
\label{eq:vc_escape}
\end{equation}

and the states are updated as

\begin{equation}
p_i^{\mathrm{act}}(t)
=
\left[1-p_i^{\mathrm{esc}}(t)\right]
p_i^{\mathrm{sus}}(t-1),
\label{eq:vc_active}
\end{equation}

\begin{equation}
p_i^{\mathrm{sus}}(t)
=
p_i^{\mathrm{esc}}(t)
p_i^{\mathrm{sus}}(t-1).
\label{eq:vc_susceptible}
\end{equation}

Iterations stop when $\max_i p_i^{\mathrm{act}}(t)<10^{-10}$ or after 10,000 steps, whichever occurs first. Viral Centrality for seed $s$, excluding the seed itself, is

\begin{equation}
VC_s(H)
=
\sum_{i=1}^{N}
\left[
1-\lim_{t\rightarrow\infty}p_i^{\mathrm{sus}}(t)
\right]
-1.
\label{eq:viral_centrality}
\end{equation}

A larger $VC_s(H)$ represents greater normalized multistep reach through the estimated directed topology. Because outgoing rows are normalized, the measure reflects relative edge allocation and downstream path structure rather than absolute outgoing connectedness. Accordingly, Viral Centrality complements NET as a normalized multistep network-position measure on the same directed topology.

\subsection{Dynamic and parameter sensitivity analyses}
\label{subsec:dynamic_sensitivity}

The full-sample analysis estimates a pooled connectedness network from the entire sample, while the intensity of housing-market connectedness may vary over time. We therefore complement the full-sample network analysis with a rolling-window analysis and sensitivity checks for the forecast horizon and PageRank damping parameter.

The rolling analysis uses a fixed window of

\begin{equation}
W=52,
\label{eq:rolling_window}
\end{equation}

corresponding approximately to one year of weekly observations. For each endpoint $t$, the estimation sample is

\begin{equation}
\mathcal{W}_t
=
\left\{
\mathbf{r}_{t-W+1},\ldots,\mathbf{r}_t
\right\}.
\label{eq:rolling_sample}
\end{equation}

Within each rolling window, the VAR and the corresponding GFEVD are re-estimated using the baseline lag specification and the fixed forecast horizon $H=10$. The rolling Total Connectedness Index is

\begin{equation}
TCI_t
=
TCI_{\mathcal{W}_t}(H=10).
\label{eq:rolling_tci}
\end{equation}

The resulting series is used to trace broad temporal changes in the overall intensity of housing-return connectedness. The 52-week window is retained to provide an approximately annual, relatively high-resolution temporal view, not because it is treated as a statistically preferred window length. With 25 variables, the corresponding VAR(1) has only 51 effective response--lag observations per rolling window, so the estimated TCI may be sensitive to finite-sample noise and to the chosen temporal resolution. Accordingly, the rolling series is interpreted as a descriptive system-level temporal profile. Window-length comparisons and the recursive TVP--VAR benchmark~\citep{antonakakis2020refined} reported in Supplementary Section~S2 assess whether the broad pattern of rising and falling connectedness is specific to the 52-week specification, while district-level temporal persistence is assessed separately in Stage~4.

The baseline network analysis uses $H=10$. To assess whether the estimated network positions depend materially on the forecast horizon, the GFEVD network and the corresponding NET, PageRank, and Viral Centrality rankings are additionally evaluated at

\begin{equation}
H\in\{5,10,15\}.
\label{eq:horizon_sensitivity}
\end{equation}

Additional numerical long-horizon benchmarks at $H=52$ and $H=260$ are reported in the Supplementary Material. Because alternative horizons recompute the GFEVD itself, both edge weights and pairwise net directions may change with $H$.

The baseline PageRank specification uses $\alpha=0.85$. Sensitivity to the damping parameter is assessed using

\begin{equation}
\alpha\in\{0.75,0.85,0.95\}.
\label{eq:pagerank_damping_sensitivity}
\end{equation}

Alternative damping values recompute the PageRank vector on the baseline transition matrix at $H=10$ and therefore leave the underlying pairwise net-connectedness network unchanged. Rank stability across alternative specifications is summarized using the Spearman rank correlation,

\begin{equation}
\rho_S(x,y)
=
1-
\frac{
6\sum_{i=1}^{N}d_i^2
}{
N(N^2-1)
},
\label{eq:spearman_rank}
\end{equation}

where $d_i$ denotes the difference between the rank of district $i$ under two alternative specifications. Tied values receive average ranks, and $\rho_S$ is computed as the Pearson correlation between the resulting rank vectors, which reduces to Eq.~\eqref{eq:spearman_rank} in the absence of ties. These analyses are descriptive sensitivity checks of the estimated network positions. Detailed rolling-window, horizon, and damping results are reported in the Supplementary Material.

\subsection{Diagnostic assessment of network-position interpretations}
\label{subsec:diagnostic_framework}

After constructing NET, PageRank, and Viral Centrality, the diagnostics quantify structural alignment and characterize condition-specific changes in network positions across all districts. The primary prominence cutoff is $k=3$, which defines a small set of prominent positions without making the assessment depend on an exact first-place ranking. Stage 1 candidate sets for $k=1$ and $k=5$ are reported as cutoff-sensitivity checks, whereas the candidate-level assessment in the subsequent stages uses $k=3$. The cutoff does not determine whether a network position is economically meaningful; positions outside the prominence set remain part of the all-node characterization.

Separately from the Stage~4 split-sample procedure, full-sample residual bootstraps are used to characterize rank uncertainty and edge-direction reproducibility around the fitted baseline network. Residual vectors from the full-sample no-intercept VAR(1) are jointly resampled in moving blocks and used to recursively generate return paths without mean-preserving recentering. Full-sample rank distributions use 2,000 replications with the 12-week reference block, with 6- and 24-week alternatives reported in Supplementary Section~S3.2. Edge-direction stability and the descriptive 90\% backbone use 10,000 replications with a three-week block selected by the automatic procedure of \citet{politis2004automatic} with the correction of \citet{patton2009correction}. The data-selected three-week block is used for full-sample edge-direction stability, whereas Stage~4 retains a common fixed block length across the early and late subsamples so that their candidate-level reproduction frequencies are evaluated under the same resampling scale. The full-sample bootstrap characterizes uncertainty around the fitted full-sample network, whereas Stage~4 preserves each subsample mean because its purpose is to assess within-period reproducibility conditional on the observed early- and late-period levels.

First, common-component sensitivity is assessed using three mean-preserving transformations. Cross-sectional-mean removal replaces $r_{i,t}$ by $r_{i,t}-(m_t-\bar m)$, where $m_t=N^{-1}\sum_{j=1}^{N}r_{j,t}$ and $\bar m$ is its sample average. PC1 removal subtracts the rank-one component obtained from the first principal-component score and loading of the centered return panel while preserving the district means. Dynamic PC1 residualization regresses the retained district returns on an intercept, the contemporaneous centered-panel PC1 score $f_t$, and its first $q$ lags; pooled BIC over $q\in\{0,\ldots,4\}$ selects $q=2$, and the residuals plus the retained-sample district means form the adjusted panel. Values and ranks are recorded for every district--measure pair, with NET signs additionally tracked for all districts. PageRank and Viral Centrality candidates require top-$k$ membership across the raw and all three adjusted systems. NET transmitter candidates require a positive upper-tail top-$k$ position across all four systems, whereas receiver candidates require a negative lower-tail top-$k$ position across all four systems. These transformations are used as comparative sensitivity representations rather than as separately identified structural systems. Because the mean-preserving adjustments constrain variation along the common direction targeted by each treatment, changes in position are interpreted as sensitivity to common movement, which may itself be economically relevant, rather than as identification of a local-shock network. Accordingly, each mean-preserving adjusted panel lies in a 24-dimensional affine subspace; the corresponding rank and stability diagnostics are reported in Supplementary Section~S4.1.

Second, structural alignment is quantified for all districts by comparing PageRank with indegree and incoming strength, and Viral Centrality with outdegree and outgoing strength. Proposition~\ref{prop:pagerank_indegree_boundary} provides a sufficient monotone-allocation condition for exact PageRank--indegree correspondence, while the controlled simulation reported in Supplementary Section~S5.2 examines how alignment changes outside this sufficient case under alternative weighted topologies. Stage~2 therefore quantifies empirical structural reproduction rather than testing whether PageRank or Viral Centrality is mathematically reducible to degree. Uniform-weight counterparts are constructed on the same directed topology by assigning unit weight to every positive edge before recomputing PageRank and Viral Centrality, and are reported descriptively. Structural rank correlations and all-node rank deviations use average ranks for ties. The rank-deviation summaries reported in Table~\ref{tab:structural_alignment} are calculated relative to the corresponding degree ranking: indegree for PageRank and outdegree for Viral Centrality. Candidate-level top-$k$ membership, however, is defined using competition ranking, so that all nodes tied at a position with competition rank no greater than $k$ are treated as belonging to the top-$k$ set. These all-node comparisons quantify the extent to which recursive or multistep rankings depart from simpler local network hierarchies. For a primary $k=3$ PageRank or Viral Centrality candidate identified in Stage 1, top-three prominence is classified as structurally aligned when it is reproduced by either the relevant degree or strength hierarchy. This classification concerns the source of candidate-level prominence rather than the mathematical meaning of the centrality measure or the full ordering it induces. Candidates whose top-three prominence is not reproduced by either corresponding degree or strength hierarchy are additionally assessed in Stage 3 for specification sensitivity. Stage 2 is not applied to NET.

Third, specification sensitivity is assessed across VAR(1), VAR(2), VAR(4), and ridge VAR(4). Rank correlations across all districts are reported for NET, PageRank, and Viral Centrality, while the ranks of the NET candidates from Stage 1 and of any PageRank or Viral Centrality candidates carried forward from Stage 2 are tracked explicitly across the four specifications. The ridge VAR is estimated after scaling each training series by its training-sample standard deviation without demeaning, thereby preserving the no-intercept specification, and the coefficients are transformed back to the original units. The penalty multiplier is selected by five-fold chronological blocked cross-validation using mean one-step-ahead squared forecast error standardized by the training-sample standard deviations. A candidate whose prominence changes across these specifications remains part of the network characterization but is described as specification conditional. Candidates whose corresponding prominence remains stable across Stage 3 proceed to the temporal-persistence assessment. This specification comparison concerns the full-sample networks; it does not imply that split-sample ranks or bootstrap reproduction frequencies are invariant to the lag specification.

Fourth, observed temporal persistence and bootstrap reproducibility are assessed separately under the BIC-selected reference VAR(1) using a fixed 50/50 chronological split. The no-intercept network at $H=10$ is estimated separately in the early and late subsamples, and the early--late rank association and changes in relative position summarize temporal reordering of the network. Candidate-level bootstrap reproducibility is assessed using 2,000 residual moving-block bootstrap replications with a 12-week block length. The 12-week block length is retained as a fixed primary resampling specification for Stage~4 so that the early and late subsamples are evaluated under the same resampling scale rather than under separately data-selected block lengths. Bootstrap samples are generated separately within the early and late subsamples using a mean-preserving residual moving-block procedure. Within each half, the district-specific sample means are removed, a no-intercept VAR(1) is fitted to the resulting deviations, and centered residual vectors are jointly resampled in moving blocks. The recursively generated deviations are recentered before the corresponding half-sample means are restored and the no-intercept VAR--GFEVD network is re-estimated. Recursion is initialized with the first observed deviation vector, and no burn-in is used. Let $P_E$ and $P_L$ denote the proportions of bootstrap replications in which the candidate reproduces its corresponding prominence state within the early and late subsample models, respectively. For NET, the relevant state is a positive upper-tail top-three position for a transmitter candidate and a negative lower-tail top-three position for a receiver candidate. The observed early--late comparison and the bootstrap reproduction frequencies are reported separately, with $P_E$ and $P_L$ treated as descriptive measures of sampling reproducibility rather than as pass--fail criteria. These quantities describe reproducibility conditional on the mean-preserving VAR(1) bootstrap generators fitted within the two subsamples and the 12-week block-resampling scheme; they do not correct residual-model misspecification or establish invariance to alternative lag orders, split dates, or forecasting settings.

The diagnostics characterize position changes and structural alignment across all districts, while the sequential candidate-level assessment traces whether prominent positions are structurally reproduced or remain conditional on the examined common-movement, dynamic-specification, and sample-timing choices.

Algorithm~\ref{alg:empirical_sequence} summarizes the computational workflow used to map the pairwise-netted network into complementary node-level positions and to evaluate their structural alignment and empirical sensitivity.

\begin{algorithm}[!htbp]
\caption{Summary of the empirical computation}
\label{alg:empirical_sequence}
\begin{algorithmic}[1]
\State Estimate the VAR--GFEVD network and construct the directed pairwise net-connectedness matrix.
\State Compute NET, PageRank, and Viral Centrality for all districts.
\State Record the full set of node values and rankings.
\State Characterize all-node common-component sensitivity and identify the primary $k=3$ candidates.
\State Quantify all-node structural alignment and document dynamic-specification, observed temporal, and bootstrap-reproducibility profiles for applicable candidates.
\State Report the all-node characterization and candidate-specific diagnostic profiles.
\end{algorithmic}
\end{algorithm}

\section{Results}
\label{sec:results}

\subsection{Baseline connectedness and node-level network positions}
\label{subsec:baseline_results}

The unadjusted Seoul VAR--GFEVD system at $H=10$ yields a TCI of 81.98\%. Songpa has TO of 201.64, FROM of 69.70, and NET of 131.95, whereas Geumcheon has NET of $-54.81$. Songpa therefore occupies the highest positive NET position in the full-sample network, while Geumcheon lies at the opposite end of the NET ordering. Gangnam is close to directional balance, with NET of 0.11, although its TO and FROM values of 77.71 and 77.60 show that a near-zero balance does not imply weak connectedness.

\begin{figure}[!htbp]
    \centering
    \includegraphics[width=0.72\textwidth]{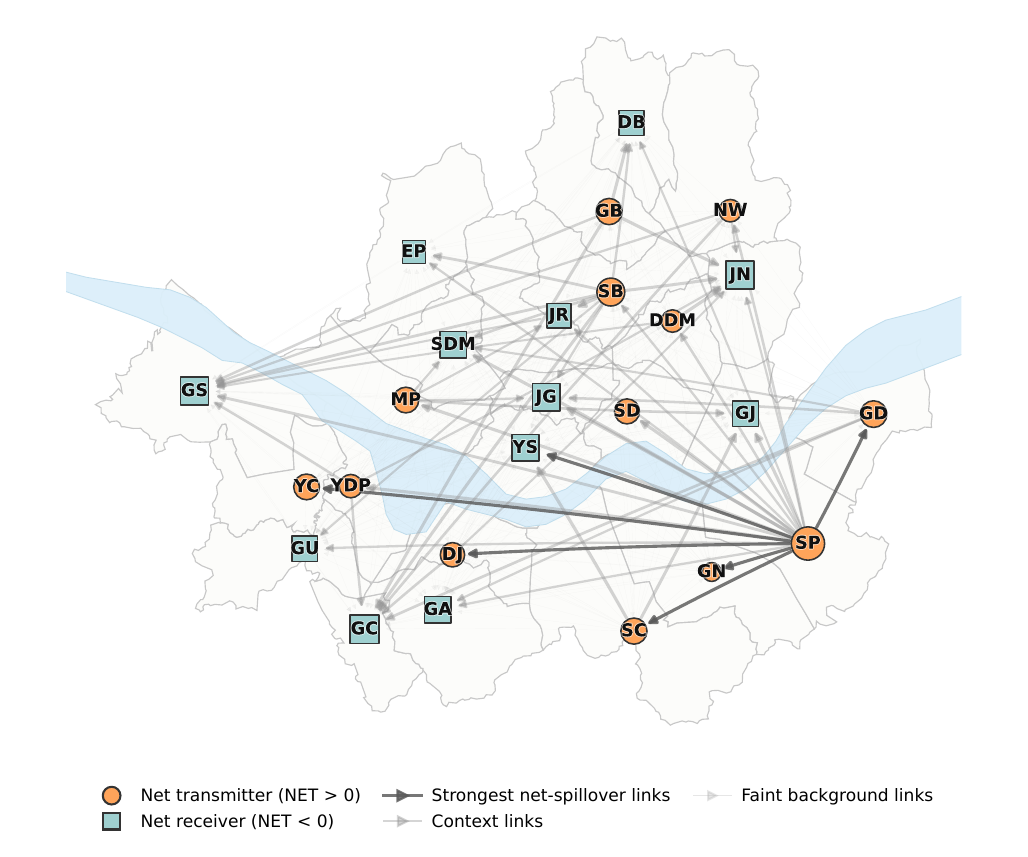}
    \caption{Directed pairwise net-connectedness network among Seoul's 25 districts at $H=10$. Circular and square nodes indicate positive and negative NET values, respectively. Node size reflects the absolute magnitude of NET, and edge width increases with $A^{+}_{ij}(H)$. All 300 positive pairwise net links are displayed, with the six strongest links emphasized, the next 50 shown as context links, and the remaining 244 shown as faint background links. Arrows indicate estimated pairwise net direction.}    \label{fig:seoul_net_network}
\end{figure}

Figure~\ref{fig:seoul_net_network} presents the directed pairwise net-connectedness network constructed from $\mathbf{A}^{+}(H)$. Songpa has positive pairwise net-connectedness differences with all other 24 districts, so its high NET is distributed across the network rather than generated by a single dominant edge. Geumcheon is the only sink in the baseline pairwise net network, with no positive outgoing link after bilateral netting. The network contains 97 cyclic triads among the 2,300 possible district triads, corresponding to a Kendall--Babington Smith consistency coefficient of $\eta_{\mathrm{KBS}}=0.851$. The baseline network therefore exhibits substantial, but incomplete, tournament consistency. Negative NET values, such as that of Geumcheon, describe net-receiver positions within the estimated connectedness network rather than the overall economic importance of the corresponding housing markets.

Using 10,000 full-sample residual-bootstrap replications with the automatically selected three-week block, the median direction-reproduction share is 0.848, while Songpa's positive NET sign and Geumcheon's negative NET sign are reproduced in 100.0\% and 99.6\% of replications, respectively. Restricting the network to the 122 links whose direction is reproduced in at least 90\% of the residual-bootstrap replications, used as a descriptive stability filter rather than a network selected by formal significance testing, nevertheless leaves the broad node-level ordering largely unchanged, with rank correlations of 0.970 for NET, 0.934 for PageRank, and 0.840 for Viral Centrality. The subsequent diagnostics therefore focus on node-level positions rather than interpreting individual arrows as independently established economic links.

\begin{figure}[!htbp]
    \centering

    \begin{minipage}[t]{0.48\textwidth}
        \centering
        \includegraphics[width=\linewidth]{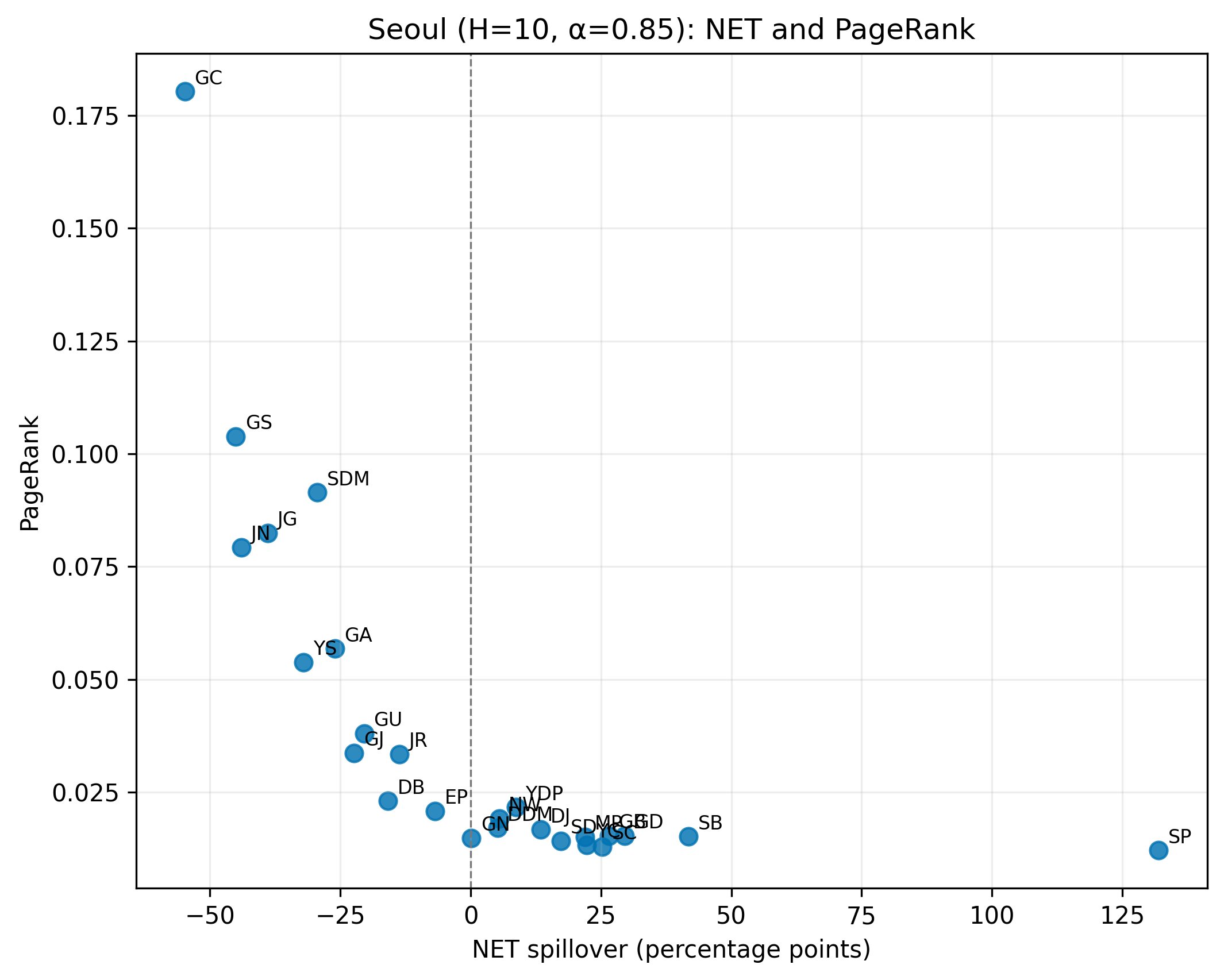}

        \small (a) NET and PageRank
    \end{minipage}
    \hfill
    \begin{minipage}[t]{0.48\textwidth}
        \centering
        \includegraphics[width=\linewidth]{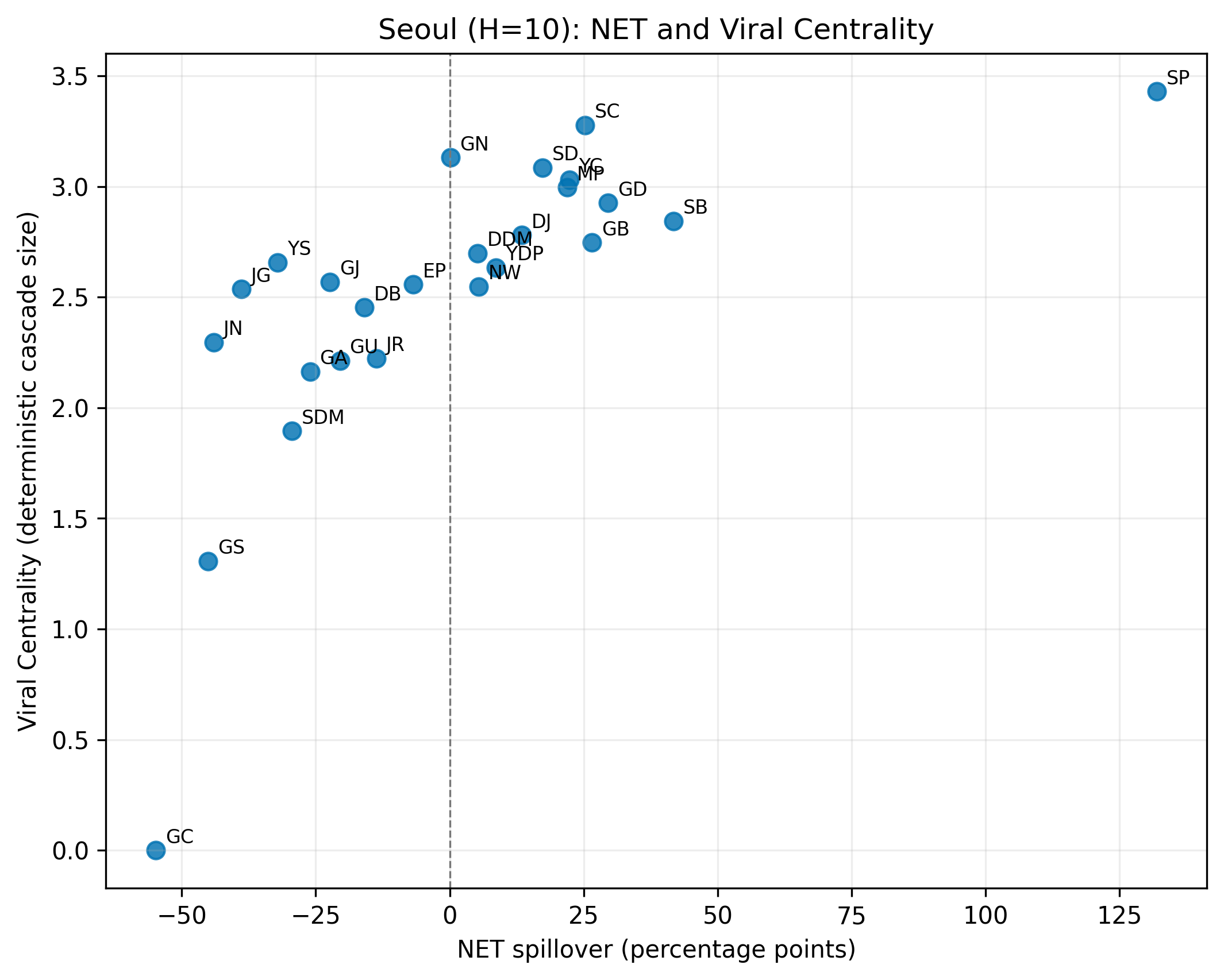}

        \small (b) NET and Viral Centrality
    \end{minipage}

    \caption{NET, PageRank, and Viral Centrality across Seoul's 25 districts at $H=10$. Panel (a) compares NET with PageRank at $\alpha=0.85$, and panel (b) compares NET with Viral Centrality. The dashed vertical line marks $NET=0$.}
    \label{fig:seoul_recursive_comparison}
\end{figure}

Figure~\ref{fig:seoul_recursive_comparison} shows contrasting district positions across the three node-level measures. Geumcheon has the highest PageRank value, 0.1803, followed by Gangseo, Seodaemun, and Jung, all of which occupy net-receiver positions. Songpa has the lowest PageRank value, 0.0121. By contrast, Songpa has the highest Viral Centrality value, 3.430, followed by Seocho at 3.278 and Gangnam at 3.132, whereas Geumcheon's Viral Centrality is zero because it is a sink in the pairwise net-connectedness network.

Gangnam is particularly informative because it ranks third in Viral Centrality despite a NET value of only 0.11. It sends positive net links to 18 of the other 24 districts, with outgoing and incoming net-connectedness strengths of approximately 0.143 and 0.141, respectively, while most of its incoming strength is associated with its link from Songpa. Its near-zero NET therefore combines a broad outgoing pattern with a concentrated incoming component, illustrating how a similar net balance can conceal substantially different directed topology. These contrasts motivate extending scalar NET with recursive and multistep positions on the same netted topology while leaving their structural distinctiveness to the quantitative diagnostics below.

\begin{figure}[!htbp]
    \centering
    \includegraphics[width=\textwidth]{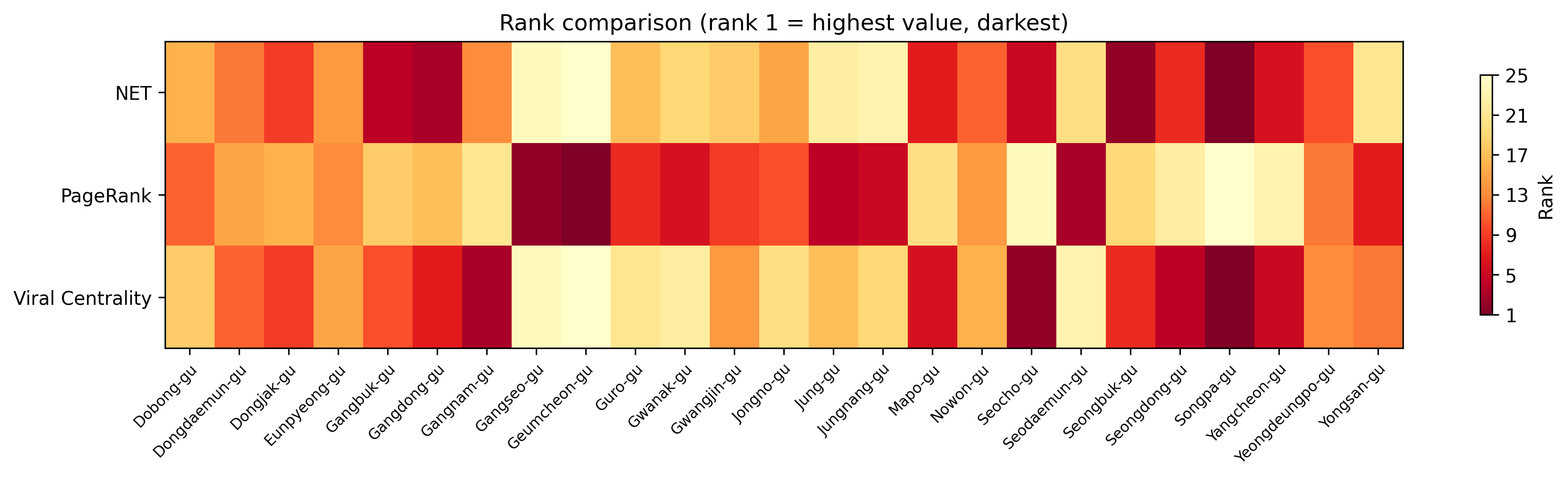}
    \caption{District rankings under NET, PageRank, and Viral Centrality at $H=10$. Districts are arranged alphabetically and displayed using their full names to facilitate consistent identification across the three rankings. PageRank is evaluated at $\alpha=0.85$. Darker cells indicate higher positions, with rank 1 denoting the largest value for each measure.}
    \label{fig:seoul_rank_comparison}
\end{figure}

Figure~\ref{fig:seoul_rank_comparison} further summarizes the district rankings under the three node-level measures. Their Spearman correlations are $-0.916$ between NET and PageRank, 0.826 between NET and VC, and $-0.937$ between PageRank and VC. The measures therefore produce strongly related but nonidentical orderings. These correlations quantify strong but incomplete agreement across the three network-position measures, but do not by themselves establish ordering information beyond simpler local network structure. That question is assessed directly in Stage~2.

\subsection{Stage 1: All-node common-component sensitivity}
\label{subsec:direct_results}

Common citywide movement is quantitatively important for the network measured from the raw district panel. Cross-sectional-mean removal, PC1 removal, and dynamic PC1 residualization reduce the TCI from 81.98\% to 51.32\%, 52.00\%, and 50.60\%, respectively. Relative to the raw system, these are reductions of 37.4\%, 36.6\%, and 38.3\%. The same treatments reverse 97, 97, and 98 of the 300 pairwise net directions, so roughly one third of the bilateral orientation changes as common movement is treated differently. The first principal component accounts for 69.5\% of centered return variation, and the corresponding rank correlations with the raw system also decline substantially. These comparisons do not identify the removed component as a structural common shock or as economically irrelevant; they show that both the magnitude and the directional organization of raw connectedness are materially conditional on citywide common movement.

\begin{table}[!htbp]
\centering
\caption{Common-component sensitivity of Seoul connectedness and rankings}
\label{tab:factor_conditional_results}
\small
\setlength{\tabcolsep}{4pt}

\begin{tabular}{@{}lrrrrr@{}}
\toprule
Treatment
& TCI (\%)
& Reversals
& $\rho_{\mathrm{NET}}$
& $\rho_{\mathrm{PR}}$
& $\rho_{\mathrm{VC}}$ \\
\midrule

Raw system
& 81.98 & 0 & 1.000 & 1.000 & 1.000 \\

Cross-sectional mean removed
& 51.32 & 97 & 0.535 & 0.782 & 0.645 \\

PC1 removed
& 52.00 & 97 & 0.544 & 0.648 & 0.499 \\

Dynamic PC1 residualization
& 50.60 & 98 & 0.412 & 0.555 & 0.432 \\

\bottomrule
\end{tabular}

\vspace{2pt}
\begin{minipage}{0.95\linewidth}
\footnotesize
\raggedright
\sloppy
\textit{Notes}. Rank correlations are Spearman correlations with the corresponding raw-system ranking across all 25 districts. Direction reversals are counted among the 300 district pairs. The raw, cross-sectional-mean, and PC1 systems use 743 return observations. Dynamic PC1 residualization uses 741 observations after pooled BIC selects $q=2$. The main system-level sensitivity patterns and the highlighted Songpa and Gangdong positions are qualitatively unchanged when the other systems are re-estimated over the same 741-observation period; matched-period results are summarized in Supplementary Section~S4.1.
\end{minipage}
\end{table}

The all-node results show that this reordering is heterogeneous. Mapo retains NET rank 7 under all four representations, demonstrating that a position outside the primary prominence set can nevertheless be stable. Geumcheon retains a negative NET value throughout, with receiver-side ranks of 1, 4, 2, and 1. Its negative NET orientation is therefore persistent even though its exact lower-tail prominence varies. In contrast, Gangdong moves from raw NET rank 3 to ranks 25, 25, and 24 after the three adjustments and changes from a positive to a negative NET value. Its prominent raw-system position is therefore strongly conditional on the treatment of citywide common movement.

Common-component treatment can also reveal positions that are less visible in the raw system. Gangnam's NET rank rises from 13th to 5th under both cross-sectional-mean removal and PC1 removal, whereas Geumcheon's first-place PageRank position falls to 10th under the same two treatments. Common-component adjustment can therefore attenuate some prominent positions while making other relative positions more visible.

The prominence cutoff is applied only after this all-node characterization. Under the primary $k=3$ rule, Stage 1 identifies Songpa--NET and Seongbuk--NET as transmitter-side candidates and Songpa--VC and Seocho--VC as VC candidates. No PageRank candidate satisfies the top-three criterion, and no NET receiver candidate satisfies the negative lower-tail $k=3$ criterion across all four representations.

The Stage 1 candidate sets also vary with the prominence cutoff. At $k=1$, Songpa is the only NET transmitter candidate, and no NET receiver, PageRank, or Viral Centrality candidate is identified. At $k=5$, Songpa, Seongbuk, and Seocho are NET transmitter candidates, Geumcheon is a NET receiver candidate, no PageRank candidate is identified, and the Viral Centrality candidate set remains Songpa and Seocho. The subsequent candidate-level assessment uses the primary cutoff $k=3$; $k=1$ and $k=5$ are reported only as Stage 1 cutoff-sensitivity checks.

\subsection{Stage 2: Structural alignment of PageRank and Viral Centrality}
\label{subsec:multidimensional_roles}

Stage 2 quantifies how closely the PageRank and Viral Centrality rankings align with their corresponding degree and strength hierarchies and identifies node-level departures from those local orderings. For the Seoul pairwise-net matrix, $\psi_G=0.7090$, so the NET-equivalent gradient component accounts for 70.9\% of the squared Frobenius norm and the NET-orthogonal residual accounts for 29.1\%. The decomposition therefore indicates that much of the pairwise-net structure is represented by the global ordering encoded by NET, while a substantial residual remains outside that scalar hierarchy. The corresponding shares in the $N=25$ transitive uniform-weight benchmark are 69.3\% and 30.7\%, respectively. By Corollary~\ref{cor:unit_weight_gradient_share}, assigning unit weights to Seoul's observed orientation gives $\psi_G=0.851\times0.6933=0.5900$. The observed weighted gradient share is therefore higher than the 59.0\% orientation-only unit-weight value, which is consistent with edge weights reinforcing the NET-related gradient component. These comparisons are descriptive and do not identify the mechanism generating that alignment. Across 50,000 controlled weighted-tournament simulations, the transitive uniform-weight case provides an exact benchmark, whereas alignment outside that benchmark is not determined by tournament consistency alone. Greater edge-weight heterogeneity generally weakens degree--centrality alignment, while the detailed arrangement of directions and weights also affects the resulting rankings.

Among the 245 simulated tournaments with consistency and positive edge-weight variation close to the Seoul values, the PageRank--indegree correlation has a median of 0.833, a descriptive 2.5th--97.5th percentile range of 0.654--0.952, and a maximum of 0.967, compared with 0.987 in Seoul. The observed Viral Centrality--outdegree correlation of 0.935 lies within the corresponding simulated range of 0.654--0.959. Because simulation weights are generated independently of orientation, this comparison shows that tournament consistency and marginal edge-weight dispersion alone do not account for Seoul's especially close PageRank--indegree ordering. It does not identify the additional structural feature responsible for that alignment. Table~\ref{tab:structural_alignment} reports the all-node empirical comparisons and the assessment of the primary $k=3$ candidates.

\begin{table}[!htbp]
\centering
\caption{Structural alignment diagnostics for PageRank and Viral Centrality}
\label{tab:structural_alignment}
\small
\setlength{\tabcolsep}{4pt}
\renewcommand{\arraystretch}{1.08}

\begin{tabular}{@{}lccccc@{}}
\toprule
\multicolumn{6}{l}{\textit{Panel A. All-node structural comparisons}} \\
\midrule
Measure
& Degree $\rho$
& Strength $\rho$
& Uniform-weight $\rho$
& Mean $|d_D|$
& Max $|d_D|$ \\
\midrule

PageRank
& 0.987
& 0.971
& 0.931
& 0.80
& 3.5 \\

Viral Centrality
& 0.935
& 0.812
& 0.905
& 1.92
& 6.5 \\

\midrule
\multicolumn{6}{l}{\textit{Panel B. Primary $k=3$ candidate assessment}} \\
\midrule
Candidate
& Degree (rank)
& Strength (rank)
& \multicolumn{3}{c}{Structural evidence} \\
\midrule

Songpa--VC
& 24 (1)
& 1.319 (1)
& \multicolumn{3}{c}{Aligned with degree and strength} \\

Seocho--VC
& 22 (2)
& 0.351 (5)
& \multicolumn{3}{c}{Aligned with degree} \\

\bottomrule
\end{tabular}

\vspace{2pt}
\begin{minipage}{0.95\linewidth}
\footnotesize
\raggedright
\sloppy
\textit{Notes}. Panel A reports Spearman rank correlations across all 25 districts using average ranks for ties. For PageRank, the degree and strength comparators are indegree and incoming strength; for Viral Centrality, they are outdegree and outgoing strength. Let $d_D$ denote the difference between the rank of the centrality measure and the rank of its corresponding degree measure, using indegree for PageRank and outdegree for Viral Centrality. Mean and maximum absolute rank deviations are reported. The uniform-weight $\rho$ is the Spearman correlation between each baseline weighted-centrality ranking and its counterpart obtained by assigning unit weight to every positive edge on the same directed topology and recomputing the same centrality measure. Panel B reports the primary $k=3$ VC candidates evaluated in Stage 2. The ranks shown in Panel B use competition ranking, under which all nodes with competition rank no greater than $k$ are included in the top-$k$ set. A candidate whose top-three prominence is reproduced by either corresponding local hierarchy is characterized as structurally aligned. Candidates whose prominence is reproduced by neither hierarchy receive additional Stage 3 specification-sensitivity assessment. No PageRank candidate is evaluated in Stage 2, and Stage 2 is not applicable to NET.
\end{minipage}
\end{table}

The all-node results indicate strong global alignment between the recursive or multistep measures and their corresponding local hierarchies. Relative to the corresponding degree ranking, the mean absolute rank deviation is 0.80 for PageRank and 1.92 for Viral Centrality, with maximum deviations of 3.5 and 6.5 ranks, respectively. Thus, strong global alignment does not imply mathematical equivalence and coexists with heterogeneous node-level departures in relative ordering.

For both primary VC candidates, top-three prominence is reproduced by at least one corresponding local hierarchy. Songpa's position is aligned with both outdegree and outgoing strength, whereas Seocho's position is aligned with outdegree but not outgoing strength. Their prominence is therefore characterized as structurally aligned under the Stage 2 rule, although this classification does not imply that all path or weight information in VC is redundant. Under the sequential candidate-level procedure, the later dynamic and temporal diagnostics were not applied to these two VC candidates. No PageRank candidate is assessed in Stage 2, and Stage 2 is not applicable to NET. Stage 3 evaluates the dynamic-specification sensitivity of Songpa--NET and Seongbuk--NET.

Geumcheon and Gangnam provide contrasting all-node examples. Geumcheon's first-place PageRank position is reproduced by both indegree and incoming strength, illustrating close structural alignment. By contrast, Gangnam ranks third in Viral Centrality while its average-rank outdegree position is 5.5 and its outgoing-strength rank is 13; under uniform edge weights, its Viral Centrality rank falls to eleventh. This difference shows that edge weights and downstream paths can alter individual node positions even when the overall centrality hierarchy remains strongly aligned with simpler local structure. Gangnam does not enter the candidate-level assessment because its Viral Centrality prominence is conditional on the common-component treatments. Additional residual-bootstrap rank distributions are reported in Supplementary Section~S3.

\subsection{Stage 3: Specification sensitivity}
\label{subsec:robustness}

Stage 3 evaluates the dynamic-specification sensitivity of Songpa--NET and Seongbuk--NET across OLS VAR(1), OLS VAR(2), OLS VAR(4), and ridge VAR(4). Table~\ref{tab:seoul_robustness} reports the system-level TCI, rank correlations, and pairwise direction reversals relative to the OLS VAR(1) baseline. Among these specifications, VAR(4) reduces the number of equation-level Ljung--Box rejections~\citep{ljung1978measure} through lag 12 from 22 of 25 under VAR(1) to 6 of 25, although joint residual whiteness remains rejected. It is therefore used as a lower-residual-autocorrelation benchmark rather than as a replacement reference model. The comparisons indicate appreciable specification sensitivity in the broader district ordering.

\begin{table}[!htbp]
\centering
\caption{Primary dynamic-specification comparisons}
\label{tab:seoul_robustness}
\small
\setlength{\tabcolsep}{4pt}

\begin{tabular}{@{}lrrrrr@{}}
\toprule
Comparison
& TCI (\%)
& $\rho_{\mathrm{NET}}$
& $\rho_{\mathrm{PR}}$
& $\rho_{\mathrm{VC}}$
& Reversals \\
\midrule

VAR(2) vs.\ VAR(1)
& 84.17 & 0.840 & 0.838 & 0.807 & 35 \\

VAR(4) vs.\ VAR(1)
& 85.37 & 0.814 & 0.779 & 0.745 & 51 \\

Ridge VAR(4) vs.\ VAR(1)
& 84.80 & 0.834 & 0.872 & 0.849 & 39 \\

\bottomrule
\end{tabular}

\vspace{2pt}
\begin{minipage}{0.95\linewidth}
\footnotesize
\raggedright
\sloppy
\textit{Notes}. TCI denotes the value obtained under the alternative specification. Rank correlations and direction reversals are calculated relative to the OLS VAR(1) baseline. Reversals are counted among the 300 district pairs.
\end{minipage}
\end{table}

The PageRank and Viral Centrality correlations are reported as all-node diagnostic summaries and do not alter the candidate set defined by the preceding stages.

Songpa retains a positive NET value and ranks first under all four specifications. Seongbuk also retains a positive NET value, but its ranks are 2, 4, 5, and 7 under VAR(1), VAR(2), VAR(4), and ridge VAR(4), respectively. Seongbuk therefore retains its positive NET orientation, while its top-three prominence is conditional on the dynamic specification. Under the primary $k=3$ assessment, Songpa--NET remains prominent across the four dynamic specifications and is the candidate evaluated in the Stage 4 temporal assessment.
Songpa's first-place NET rank is therefore retained in the specification with substantially lower equation-level residual autocorrelation. At the same time, the VAR(4)-to-VAR(1) rank correlations of 0.814, 0.779, and 0.745 for NET, PageRank, and Viral Centrality show that finer all-node ordering remains specification-dependent.

Additional sensitivity checks provide a broader context for these specification results. Relative to $H=10$, the NET, PageRank, and Viral Centrality rank correlations are 0.985, 0.944, and 0.935 at $H=5$ and 0.998, 0.986, and 0.989 at $H=15$. At the numerical long-horizon benchmark $H=260$, the corresponding correlations remain 0.995, 0.960, and 0.926, although 17 pairwise net directions differ from the $H=10$ network. PageRank is nearly invariant to the examined damping values, with rank correlations of 1.000 between $\alpha=0.75$ and 0.85 and 0.998 between $\alpha=0.85$ and 0.95. Excluding the three returns spanning 14-day intervals yields a TCI of 80.96\%, 28 direction reversals, and rank correlations of 0.889, 0.901, and 0.846 for NET, PageRank, and Viral Centrality. Including an intercept yields a TCI of 81.65\% and corresponding rank correlations of 0.994, 0.944, and 0.992. Across these checks, Songpa remains first in NET at every evaluated horizon, after the gap adjustment, and in the intercept-included VAR(1). Detailed results are reported in the Supplementary Material.

\subsection{Stage 4: Temporal persistence and bootstrap reproducibility}
\label{subsec:oos_persistence}

Within the BIC-selected reference VAR(1), Stage 4 separately evaluates observed split-sample persistence and bootstrap reproducibility for Songpa--NET using the fixed 50/50 chronological split. The early subsample covers May 14, 2012 to June 24, 2019 and contains 371 observations, while the late subsample covers July 1, 2019 to August 24, 2026 and contains 372 observations. At the system level, the early--late NET rank correlation is 0.351, and the two subsamples share only one of their three highest-ranked districts. The overall NET ordering therefore exhibits substantial temporal reordering.

At the candidate level, Songpa has a positive NET value and ranks first in both observed subsamples. Across 2,000 mean-preserving residual moving-block bootstrap replications with a 12-week block length, its positive top-three NET position is reproduced in 71.45\% of the early-subsample replications and 95.10\% of the late-subsample replications. Denoting these within-subsample reproduction frequencies by $P_E$ and $P_L$, respectively,
\[
P_E=0.7145,
\qquad
P_L=0.9510.
\]
The lower early-period frequency indicates weaker residual-bootstrap reproducibility around the fitted early-subsample generator than around the late-subsample generator.

\subsection{Integrated diagnostic assessment}
\label{subsec:integrated_assessment}

Table~\ref{tab:diagnostic_profiles} summarizes the common-movement, structural, dynamic-specification, and temporal evidence for the primary $k=3$ candidates separately. The table distinguishes diagnostics that were not applicable, diagnostics that were not assessed under the sequential candidate-level procedure, and evidence showing condition-dependent prominence.

\begin{table}[!htbp]
\centering
\caption{Summary of diagnostic evidence for the primary $k=3$ candidates}
\label{tab:diagnostic_profiles}
\scriptsize
\setlength{\tabcolsep}{1pt}
\renewcommand{\arraystretch}{1.15}

\begin{tabular}{@{}
L{0.11\linewidth}
L{0.14\linewidth}
L{0.14\linewidth}
L{0.15\linewidth}
L{0.20\linewidth}
L{0.16\linewidth}
@{}}
\toprule
Node--measure pair
& Common movement
& Structural alignment
& Dynamic specification
& Temporal evidence
& Interpretation \\
\midrule

Songpa--NET
& Positive top-three in the raw and all three adjusted systems
& Not applicable
& Rank 1 under all four VAR specifications
& Under reference VAR(1), rank 1 in both observed subsamples;
  mean-preserving bootstrap $P_E=0.7145$ and $P_L=0.9510$
& Positive prominence retained across the common-movement and
  dynamic-specification analyses; bootstrap reproducibility is
  higher in the late than in the early subsample model \\
  
Seongbuk--NET
& Positive top-three in the raw and all three adjusted systems
& Not applicable
& Positive NET under all four specifications; ranks 2, 4, 5, and 7
& Not assessed
& Positive orientation retained; top-three prominence is
  specification conditional \\

Songpa--VC
& Top-three in the raw and all three adjusted systems
& Top-three prominence reproduced by outdegree and outgoing strength
& Not assessed
& Not assessed
& Prominence reproduced by simpler outgoing hierarchies \\

Seocho--VC
& Top-three in the raw and all three adjusted systems
& Top-three prominence reproduced by outdegree
& Not assessed
& Not assessed
& Prominence reproduced by the outdegree hierarchy \\

\bottomrule
\end{tabular}

\vspace{2pt}
\begin{minipage}{0.95\linewidth}
\footnotesize
\raggedright
\sloppy
\textit{Notes}. The primary prominence cutoff is $k=3$. ``Not applicable'' indicates that the diagnostic does not apply to the corresponding measure. ``Not assessed'' indicates that the diagnostic was not applied under the sequential candidate-level procedure; it does not constitute evidence against the corresponding estimated network position. $P_E$ and $P_L$ denote the within-subsample bootstrap reproduction frequencies for Songpa's positive top-three NET position in the early and late subsample models, respectively. Neither a NET receiver candidate nor a PageRank candidate meets the Stage 1 prominence condition. The table summarizes candidate-level diagnostic evidence and does not exhaust the broader all-node characterization.
\end{minipage}
\end{table}

The combined diagnostic evidence reveals three distinct sources of conditionality. Songpa--NET remains prominent across the common-movement and full-sample dynamic-specification analyses. Under the reference VAR(1), it ranks first in both observed subsamples, although its bootstrap reproduction frequency is lower in the early subsample. Seongbuk retains a positive NET orientation, while its relative prominence varies across dynamic specifications. The prominent VC positions of Songpa and Seocho are reproduced by simpler outgoing hierarchies. Diagnostics not applied under the sequential procedure are reported as not assessed and are not interpreted as adverse evidence regarding the corresponding estimated positions.

These candidate-level results summarize only the prominent positions assessed under the sequential procedure. Positions outside the candidate set remain informative components of the all-node network characterization, including positions that are stable outside the top-three cutoff or that become more or less prominent under particular diagnostic conditions.

\section{Discussion}
\label{sec:discussion}

\subsection{Common movement and district-level network positions}

The common-component analysis changes the economic reading of the raw connectedness network. Across the three treatments, 36.6--38.3\% of the raw TCI is removed and 97--98 of the 300 bilateral net directions are reversed. The raw network therefore combines district-level directional dependence with a substantial component that is sensitive to citywide co-movement. This does not imply that the removed component is economically unimportant, nor do the transformations identify a structural common shock. It does imply that a raw VAR--GFEVD edge should not be read mechanically as a stable district-to-district ripple channel.

The district-level results show that this sensitivity is heterogeneous. Geumcheon retains a negative NET in all four representations, although its receiver-side rank changes. Gangdong changes from a positive to a negative NET and falls sharply in rank after adjustment, whereas Mapo retains NET rank 7 in all four representations. These cases distinguish stability of the NET sign from stability of relative rank and top-three membership. Gangnam provides a related example: its near-zero NET in the raw network coexists with substantial connectedness in both directions because broad outgoing links are approximately offset by concentrated incoming connectedness, while its prominent Viral Centrality position reflects the allocation of outgoing weights and downstream directed paths. Thus, a near-zero NET does not imply weak involvement in the network, and a prominent raw-network position need not survive alternative treatments of common movement.

Songpa and Seongbuk further illustrate the distinction between the NET sign and relative rank. Songpa remains in the positive top three across the common-component treatments and dynamic specifications and ranks first in both observed chronological subsamples. Its positive top-three position is nevertheless reproduced less frequently in the early subsample bootstrap than in the late subsample bootstrap. Seongbuk retains a positive NET across the reported specifications, but its rank moves outside the top three under some specifications. Stability of the NET sign, stability of relative rank, and bootstrap reproducibility should therefore be interpreted separately.

Prior evidence for Seoul also illustrates why the particular network summary matters. 
\citet{hwang2021analyzing} report that the overall influence of the three Gangnam-area districts is low, although their movements tend to lead the total connectedness index by a few months. In the present district-level profiles, the three districts are 
heterogeneous: Songpa occupies a comparatively stable positive NET position, 
Gangnam is close to directional balance in the raw network, and Seocho's prominent 
Viral Centrality position is structurally aligned with its outgoing degree hierarchy. 
These findings are not directly contradictory because the studies summarize different 
dimensions of connectedness and use different samples and model constructions. Rather, 
they indicate that a group-level description of geographically related housing markets 
can combine markedly different district-level network positions.

Songpa's positive NET describes an outward-tilted position in the estimated interdistrict network: after aggregation across bilateral relations, connectedness directed from Songpa to other districts exceeds connectedness directed toward it. Possible economic explanations include transaction activity, linkages with geographically or economically related submarkets, and heterogeneous district responses to broader housing-market conditions \citep{meen1999regional,clayton2010pricevolume}. The present analysis cannot distinguish among these explanations because district-level transaction activity, local fundamentals, credit conditions, and expectation measures are not included directly. Songpa is therefore interpreted as occupying a reduced-form relative position in the estimated spatial network, not as the structural origin of housing-market shocks.

\subsection{Hierarchical structure, directional feedback, and centrality alignment}

The least-squares edge-flow decomposition provides a useful bridge between the conventional NET measure and the richer pairwise network. Because the gradient component is in one-to-one correspondence with NET, it represents the part of bilateral net flow that can be organized by a single district potential. In economic terms, this is consistent with an upstream--downstream form of the housing-market ripple described in the regional literature~\citep{meen1999regional}. The NET-orthogonal residual contains directional variation that cannot be represented by that single ordering. In Seoul, the gradient and residual shares are 70.9\% and 29.1\%, respectively, which is consistent with a substantial hierarchical component without reducing the estimated network to a purely hierarchical ripple structure. This decomposition concerns weighted pairwise flow, whereas the Kendall--Babington Smith coefficient summarizes consistency of edge orientations; the two diagnostics therefore describe different aspects of the same directed network.

The close alignment of PageRank with indegree and Viral Centrality with outdegree reflects another structural feature of the Seoul network. Because its directed support forms a tournament, every node satisfies
\[
d_i^{\mathrm{in}} = N-1-d_i^{\mathrm{out}}.
\]
Incoming and outgoing degree therefore represent opposite sides of the same tournament degree hierarchy rather than independent topological dimensions. Related connectedness studies likewise show that a receiver-side position can coexist with high centrality \citep{fu2026dynamic}.

PageRank and Viral Centrality nevertheless add edge weights, recursive allocation, and directed paths to this degree hierarchy, distinguishing not only how many net directions point toward or away from a district but also how those links are weighted and where subsequent paths lead. Proposition~\ref{prop:pagerank_indegree_boundary} establishes a sufficient monotone-allocation condition for exact PageRank--indegree rank correspondence in a transitive tournament, with uniform outgoing allocation as a special case. It does not provide a general characterization of rank correspondence in weighted or cyclic tournaments. In the baseline Seoul network, 97 of the 2,300 district triads are cyclic, corresponding to $\eta_{\mathrm{KBS}}=0.851$. The strong association between PageRank and the incoming degree hierarchy, together with the close alignment between Viral Centrality and the outgoing degree hierarchy, is consistent with the pronounced transmission--reception hierarchy. The controlled simulations further show that tournament consistency alone does not determine this alignment, because edge-weight heterogeneity and the detailed arrangement of directions and weights also shape the resulting rankings. The observed alignment is therefore interpreted as a property of the realized weighted network structure rather than as mathematical equivalence of the measures or a mechanical consequence of tournament consistency.

The 245 simulated networks matched on tournament consistency and weight dispersion have lower PageRank--indegree alignment than Seoul under independent weight assignment. Independent weight assignment does not impose the rowwise monotone allocation identified in Proposition~\ref{prop:pagerank_indegree_boundary}. Together with the gradient-share comparison, this result is consistent with the testable hypothesis that dependence between edge weights and the relative positions of their endpoint nodes contributes to Seoul's unusually high alignment. Because the Seoul network contains 97 cyclic triads, however, the proposition does not apply directly to the empirical network, and this possibility is treated as a hypothesis rather than an identified explanation.

The candidate-level comparisons make this structural alignment more concrete. Songpa's prominent Viral Centrality position is also prominent under both corresponding local outgoing measures, while Seocho's prominence is reproduced by its outdegree position. Their Viral Centrality values remain interpretable summaries of normalized multistep reach, but their top-three prominence does not define a separate hierarchy from the corresponding simpler outgoing structure under the Stage 2 criterion. This conclusion concerns the organization of the observed ranking rather than the mathematical meaning or validity of Viral Centrality as a network measure.

Gangnam illustrates a node-level departure from this broader hierarchy. Its raw-network Viral Centrality position is considerably more prominent than its outgoing-strength position, and assigning uniform weight to the positive edges substantially lowers its Viral Centrality ranking. Its near-zero NET reflects an approximate balance between outgoing and incoming net-connectedness strength, whereas its high raw-network Viral Centrality reflects the allocation of its outgoing links and their downstream directed paths. Edge weights and downstream topology can therefore alter an individual district's position even when the overall ranking structure is strongly organized by the broader transmission--reception hierarchy. Because Gangnam's Viral Centrality rank declines after common-component adjustment, this departure is interpreted as a feature of the raw network rather than as a stable district characteristic.

The contemporaneous-horizon benchmark provides a related structural comparison. Without lag propagation, the pairwise orientation collapses to a transitive ordering determined by each district's sum of squared contemporaneous correlations, whereas the baseline multistep network contains cyclic directed structure. The baseline topology therefore reflects lag-dependent information that is absent from the purely contemporaneous benchmark, even though much of its recursive ordering remains aligned with simple degree structure. Detailed results for this benchmark are reported in Supplementary Section~S4.

Because PageRank and Viral Centrality are constructed from row-normalized edge allocations, with PageRank additionally incorporating dangling-node correction and teleportation, they emphasize relative allocation and network structure rather than absolute connectedness magnitude. They are therefore interpreted jointly with NET and the underlying edge weights rather than as calibrated measures of economic propagation or causal influence.

\subsection{Dynamic specification, temporal persistence, and bootstrap reproducibility}

The specification diagnostics show that district ordering is not invariant to the dynamic model. VAR(4) shows substantially lower equation-level residual autocorrelation than VAR(1) but does not eliminate joint residual dependence, while regularization changes coefficient estimation without identifying a uniquely correct dynamic representation. The alternative specifications are therefore used as full-sample sensitivity benchmarks rather than as competing claims about the true data-generating process.
Accordingly, Songpa's retained first-place NET position under VAR(4) is interpreted as robustness to a lower-residual-autocorrelation benchmark, not as evidence that residual dependence has been resolved.

This distinction is reflected in the candidate profiles. Songpa retains a positive and highly prominent NET position across the reported dynamic specifications, whereas Seongbuk remains on the positive net-transmitting side but its relative prominence changes with the specification. The contrast suggests that stability of directional orientation and stability of rank prominence are separate properties. A district may retain a consistent position on one side of the NET balance while its relative location within that side remains sensitive to the assumed dynamics.

Additional numerical and data-construction checks provide broader context for these results. Alternative forecast horizons and the irregular-interval adjustment preserve much of the baseline district ordering, while the intercept and GFEVD-normalization checks leave the broader NET ordering similar. PageRank is also relatively insensitive to the examined damping choices. Taken together, the broad ordering is more stable than finer directions and ranks across the examined settings. These checks are descriptive sensitivity analyses rather than additional candidate-level decision rules. Detailed results are reported in Supplementary Sections~S1.1 and~S4.

The rolling analyses trace how system-wide connectedness evolves over time, providing a time-varying view of the Seoul housing market that is not visible from the full-sample network alone. The 52-week rolling TCI traces broad phases of rising and falling market-wide connectedness over the sample. The same broad temporal pattern is visible in the 104-week specification and the recursive TVP--VAR benchmark, although their absolute levels and smoothness differ with temporal resolution. These system-level paths therefore characterize the broader evolution of Seoul's housing-market connectedness, while district-level temporal positions are assessed separately in Stage~4. The corresponding time-varying comparisons are reported in Supplementary Section~S2.

Temporal persistence is also candidate specific rather than system wide. Under the reference VAR(1), the chronological split substantially reshuffles the overall NET ordering, yet Songpa remains on the positive side and occupies the leading NET position in both observed subsamples. Mean-preserving residual resampling produces different levels of support in the two subsample bootstrap distributions, with the positive top-three NET position reproduced less frequently in the early subsample than in the late subsample. These results characterize temporal evidence within the reference specification and do not establish invariance of split-sample ranks or bootstrap frequencies to alternative dynamic specifications, split dates, or forecasting settings.

\subsection{Data construction, edge-level uncertainty, and scope}

Several limitations bound these interpretations. First, the common-component transformations are comparative statistical adjustments rather than identified structural shocks or standalone structural systems. Their mean-preserving construction constrains variation along the common direction targeted by each treatment, so the adjusted networks characterize sensitivity of estimated positions to common movement rather than identify its economic source.

Second, individual pairwise directions are estimated with substantial uncertainty. The residual bootstrap is therefore used to describe how frequently each estimated direction is reproduced rather than to conduct formal edge-level significance tests. Restricting the network to links whose estimated directions are reproduced in at least 90\% of replications provides a descriptive stability filter and leaves the broad node ordering qualitatively similar. Edge-direction stability and node-level positional stability address different empirical questions. This does not imply the same degree of uncertainty for node-level summaries, which aggregate information across all bilateral differences associated with a district. Under the same resampling design, the signs of the extreme NET positions are reproduced much more consistently at the node level. This distinction motivates the emphasis on district-level network positions and diagnostic profiles rather than on individual arrows as independently established transmission channels. Detailed results are reported in Supplementary Section~S3.

Third, residual-bootstrap rank distributions show that reproducibility itself differs across measures and districts. Some broad positions recur across alternative block lengths, whereas exact ranks and several intermediate positions are more variable. Rank-based interpretations should therefore emphasize persistent relative position or prominence when supported, rather than treating the full estimated ordering as fixed. The Stage~4 split-sample bootstrap uses mean-preserving VAR(1) generators fitted to deviations around the early- and late-subsample means, with a fixed 12-week block length, so its reproduction frequencies remain conditional on both the fitted dynamic specification and the resampling scheme. Detailed full-sample rank-distribution results are reported in Supplementary Section~S3.

Fourth, the Korea Real Estate Board series have undergone periodic sample adjustments and redesigns that cannot be separately corrected with the available data. Survey updating may smooth weekly changes or introduce heterogeneous reporting lags across districts. Residual resampling, irregular-interval checks, common-component treatments, and alternative dynamic specifications cannot eliminate an unknown district-specific index-updating process. This limitation concerns the mapping from the observed REB indices to latent district-level market-price movements rather than the internal calculation of the VAR--GFEVD network from the observed series. The reported network positions should therefore be interpreted within the observed index system.

Finally, the low-connectedness New York City exercise reported in Supplementary Section~S6 examines how the same computational framework behaves when connectedness is weak and rank stability is limited. It is used as a computational stress test rather than as external validation or a direct cross-city replication of the Seoul district-level ordering. Its purpose is to show that the framework need not mechanically produce sharply differentiated or persistent network positions in every application. Together, these diagnostics evaluate the empirical sensitivity and reproducibility of the estimated network positions under observed-data transformations, alternative specifications, resampling schemes, and contrasting connectedness conditions. 

More generally, the analysis characterizes reduced-form housing-return connectedness. It does not identify structural causal transmission, the economic mechanisms generating the estimated network, or predictive effects.

\section{Conclusion}
\label{sec:conclusion}

The Seoul results point to two related limits on a literal reading of housing connectedness networks. First, alternative treatments of citywide common movement reduce the raw TCI by 36.6--38.3\% and reverse roughly one third of bilateral net directions. Second, the pairwise-net matrix contains both a NET-equivalent gradient component and a NET-orthogonal residual: 70.9\% of its squared Frobenius norm is represented by the gradient and 29.1\% remains outside that global ordering. The estimated network therefore contains a strong upstream--downstream organization, but neither its aggregate magnitude nor all of its bilateral directions can be treated as a fixed map of district-to-district ripple effects.

The computational contribution is a framework that combines a NET-specific least-squares edge-flow decomposition, a sufficient structural benchmark for PageRank--indegree alignment, controlled network simulations, and empirical sensitivity diagnostics. The decomposition gives a precise interpretation of what scalar NET retains from the full pairwise-net matrix, while the PageRank result establishes rank correspondence with indegree under monotone outgoing allocation in a transitive tournament. The simulations examine how this correspondence changes with cyclic organization and weight heterogeneity, and the empirical comparisons assess alignment with degree and strength rankings. In Seoul, PageRank and Viral Centrality closely reproduce incoming and outgoing degree rankings, although edge weights and directed paths still change individual district positions. Combining these structural comparisons with common-component, dynamic-specification, and temporal analyses makes both the basis and the stability of a prominent network position explicit.

Songpa provides the clearest example of observed NET persistence: it remains prominent across the examined common-component treatments and VAR specifications and ranks first in both chronological subsamples. Its positive top-three NET position is reproduced in 71.45\% of early-subsample bootstrap replications and 95.10\% of late-subsample replications, distinguishing observed persistence from sampling reproducibility. Gangnam illustrates a different issue: its near-zero NET coexists with high raw-network Viral Centrality, but that prominence changes under common-component treatment. These findings support reporting district positions with the conditions under which they persist. Their economic interpretation remains tied to the observed price indices and fitted reduced-form models. Linking these positions to transactions, credit conditions, and local fundamentals would help explain the mechanisms behind them.

\bibliography{sn-bibliography}

\section*{Statements and Declarations}

\section*{Acknowledgments}
The work of S.~Jang was supported by a National Research Foundation of Korea (NRF) grant funded by the Korea government (No.~RS-2024-00464395).
The work of H.~Lee was supported in part by an NRF grant funded by the Korea government (MSIT) (No.~RS-2024-00408003) and by a KIAS Individual Grant (No.~AP103101) through the Center for AI and Natural Sciences at the Korea Institute for Advanced Study.

\subsection*{Competing interests}

The authors declare that they have no competing interests.

\subsection*{Ethics approval}

Not applicable.

\subsection*{Consent to participate}

Not applicable.

\subsection*{Consent for publication}

Not applicable.

\subsection*{Data availability}

The Seoul housing-price data used in this study are available from the R-ONE Real Estate Statistics Information System of the Korea Real Estate Board. The New York City property transaction data are available from the New York City Department of Finance.

\subsection*{Code availability}

The code, configuration files, and machine-readable results used to reproduce the reported analyses are available at \url{https://github.com/tjdls034-gif/housing-connectedness-replication}.

\subsection*{Author contributions}

Seo In Jang: Conceptualization, Methodology, Formal analysis, Writing--original draft.  
Seung Ho Choi: Data curation, Software, Validation, Visualization, Writing--original draft.  
Ji-Eun Shin: Investigation, Validation, Writing--review and editing.  
Hyunwoo Lee: Supervision, Methodology, Writing--review and editing.

\clearpage
\section*{Supplementary Information}

\setcounter{table}{0}
\renewcommand{\thetable}{S\arabic{table}}
\setcounter{figure}{0}
\renewcommand{\thefigure}{S\arabic{figure}}

\subsection*{S1. Stationarity, distributional diagnostics, VAR lag selection, and irregular intervals}

This section reports the detailed stationarity and distributional diagnostics underlying the baseline VAR specification, the lag-selection results, and the sensitivity analysis for the three irregular 14-day return intervals. The corresponding empirical implications are summarized in the main text.

\begin{table}[!htbp]
\centering
\caption{Augmented Dickey--Fuller and Phillips--Perron unit-root tests for the weekly housing-return series}
\label{tab:S1_stationarity}
\scriptsize
\setlength{\tabcolsep}{3.5pt}
\renewcommand{\arraystretch}{1.08}

\begin{tabular}{@{}lrrrrrr@{}}
\toprule
Series
& \shortstack{ADF\\statistic}
& \shortstack{ADF\\$p$-value}
& \shortstack{ADF\\lag}
& \shortstack{PP\\statistic}
& \shortstack{PP\\$p$-value}
& \shortstack{PP\\bandwidth} \\
\midrule
\multicolumn{7}{@{}l}{\textit{Seoul districts}} \\
Gangnam      & $-6.296$ & $<0.001$ & 1 & $-8.954$  & $<0.001$ & 20 \\
Gangdong     & $-6.005$ & $<0.001$ & 1 & $-7.634$  & $<0.001$ & 20 \\
Gangbuk      & $-3.567$ & $<0.001$ & 1 & $-5.894$  & $<0.001$ & 20 \\
Gangseo      & $-4.996$ & $<0.001$ & 1 & $-7.327$  & $<0.001$ & 20 \\
Gwanak       & $-6.652$ & $<0.001$ & 1 & $-10.303$ & $<0.001$ & 20 \\
Gwangjin     & $-6.438$ & $<0.001$ & 1 & $-9.484$  & $<0.001$ & 20 \\
Guro         & $-4.287$ & $<0.001$ & 1 & $-7.017$  & $<0.001$ & 20 \\
Geumcheon    & $-5.204$ & $<0.001$ & 1 & $-12.455$ & $<0.001$ & 20 \\
Nowon        & $-4.062$ & $<0.001$ & 1 & $-6.256$  & $<0.001$ & 20 \\
Dobong       & $-4.293$ & $<0.001$ & 1 & $-6.579$  & $<0.001$ & 20 \\
Dongdaemun   & $-4.489$ & $<0.001$ & 1 & $-7.527$  & $<0.001$ & 20 \\
Dongjak      & $-5.416$ & $<0.001$ & 1 & $-8.244$  & $<0.001$ & 20 \\
Mapo         & $-5.545$ & $<0.001$ & 1 & $-5.914$  & $<0.001$ & 20 \\
Seodaemun    & $-5.237$ & $<0.001$ & 1 & $-10.035$ & $<0.001$ & 20 \\
Seocho       & $-5.780$ & $<0.001$ & 1 & $-6.689$  & $<0.001$ & 20 \\
Seongdong    & $-6.361$ & $<0.001$ & 1 & $-7.370$  & $<0.001$ & 20 \\
Seongbuk     & $-3.581$ & $<0.001$ & 1 & $-4.187$  & $<0.001$ & 20 \\
Songpa       & $-6.264$ & $<0.001$ & 1 & $-7.240$  & $<0.001$ & 20 \\
Yangcheon    & $-6.668$ & $<0.001$ & 1 & $-9.560$  & $<0.001$ & 20 \\
Yeongdeungpo & $-4.735$ & $<0.001$ & 1 & $-6.963$  & $<0.001$ & 20 \\
Yongsan      & $-5.230$ & $<0.001$ & 1 & $-5.851$  & $<0.001$ & 20 \\
Eunpyeong    & $-3.827$ & $<0.001$ & 1 & $-5.926$  & $<0.001$ & 20 \\
Jongno       & $-3.297$ & $<0.001$ & 1 & $-4.520$  & $<0.001$ & 20 \\
Jung         & $-5.859$ & $<0.001$ & 1 & $-7.957$  & $<0.001$ & 20 \\
Jungnang     & $-6.290$ & $<0.001$ & 1 & $-13.037$ & $<0.001$ & 20 \\
\bottomrule
\end{tabular}

\vspace{2pt}
\begin{minipage}{0.95\linewidth}
\footnotesize
\raggedright
\sloppy
\textit{Notes}. ADF denotes the Augmented Dickey--Fuller test and PP denotes the Phillips--Perron $Z_{\tau}$-type test. The null hypothesis of both tests is that the weekly return series contains a unit root. Both tests are conducted without a constant or deterministic trend. ADF lag denotes the number of lagged differences, fixed at the system-wide BIC-selected VAR lag order, and PP bandwidth denotes the Bartlett-kernel Newey--West bandwidth $\lceil 12(T/100)^{1/4}\rceil=20$. The unit-root rejection remains unchanged when the ADF test includes a constant and uses series-specific AIC-selected lags up to 12; this is treated as a stationarity check rather than evidence of dynamic adequacy. The table uses the 743 weekly Seoul observations.
\end{minipage}
\end{table}

\begin{table}[!htbp]
\centering
\caption{KPSS stationarity and Jarque--Bera normality diagnostics for the weekly housing-return series}
\label{tab:S1_kpss_jb}
\scriptsize
\setlength{\tabcolsep}{4pt}
\renewcommand{\arraystretch}{1.08}

\begin{tabular}{@{}lrrrrr@{}}
\toprule
Series
& \shortstack{KPSS\\statistic}
& \shortstack{KPSS\\$p$-value}
& \shortstack{KPSS\\lag}
& \shortstack{JB\\statistic}
& \shortstack{JB\\$p$-value} \\
\midrule
Gangnam      & 0.257 & 0.100 & 16 & 836.36  & $<0.001$ \\
Gangdong     & 0.213 & 0.100 & 16 & 712.38  & $<0.001$ \\
Gangbuk      & 0.192 & 0.100 & 17 & 5420.31 & $<0.001$ \\
Gangseo      & 0.207 & 0.100 & 17 & 696.03  & $<0.001$ \\
Gwanak       & 0.208 & 0.100 & 16 & 2590.53 & $<0.001$ \\
Gwangjin     & 0.438 & 0.061 & 16 & 5328.27 & $<0.001$ \\
Guro         & 0.216 & 0.100 & 16 & 273.34  & $<0.001$ \\
Geumcheon    & 0.231 & 0.100 & 16 & 7694.99 & $<0.001$ \\
Nowon        & 0.146 & 0.100 & 17 & 7834.50 & $<0.001$ \\
Dobong       & 0.180 & 0.100 & 17 & 6317.30 & $<0.001$ \\
Dongdaemun   & 0.183 & 0.100 & 17 & 1745.48 & $<0.001$ \\
Dongjak      & 0.263 & 0.100 & 16 & 835.69  & $<0.001$ \\
Mapo         & 0.380 & 0.086 & 16 & 3966.86 & $<0.001$ \\
Seodaemun    & 0.205 & 0.100 & 16 & 2983.21 & $<0.001$ \\
Seocho       & 0.423 & 0.067 & 16 & 249.96  & $<0.001$ \\
Seongdong    & 0.549 & 0.031 & 16 & 2468.54 & $<0.001$ \\
Seongbuk     & 0.223 & 0.100 & 17 & 4814.41 & $<0.001$ \\
Songpa       & 0.440 & 0.060 & 16 & 818.21  & $<0.001$ \\
Yangcheon    & 0.433 & 0.063 & 16 & 980.71  & $<0.001$ \\
Yeongdeungpo & 0.400 & 0.077 & 16 & 628.30  & $<0.001$ \\
Yongsan      & 0.752 & 0.010 & 16 & 1409.39 & $<0.001$ \\
Eunpyeong    & 0.171 & 0.100 & 17 & 6681.62 & $<0.001$ \\
Jongno       & 0.238 & 0.100 & 17 & 2282.94 & $<0.001$ \\
Jung         & 0.328 & 0.100 & 16 & 4160.96 & $<0.001$ \\
Jungnang     & 0.174 & 0.100 & 16 & 5597.69 & $<0.001$ \\
\bottomrule
\end{tabular}

\vspace{2pt}
\begin{minipage}{0.95\linewidth}
\footnotesize
\raggedright
\sloppy
\textit{Notes}. The KPSS null hypothesis is level stationarity. The test includes a constant and uses automatic lag selection. Using unadjusted $p$-values, the stationarity null is rejected for Seongdong and Yongsan at the 5\% level. No KPSS rejection remains after Benjamini--Hochberg adjustment across the 25 district series. The Jarque--Bera null hypothesis is normality and is rejected for all 25 series before and after the same adjustment. Reported KPSS $p$-values of 0.100 are at the upper boundary returned by the test implementation.
\end{minipage}
\end{table}

\begin{table}[!htbp]
\centering
\caption{BIC-based VAR lag-order selection for the Seoul system}
\label{tab:S1_lag_selection}
\small
\setlength{\tabcolsep}{6pt}
\renewcommand{\arraystretch}{1.08}

\begin{tabular}{@{}lrrrrr@{}}
\toprule
Application & 1 & 2 & 3 & 4 & 5 \\
\midrule
Seoul
& $\mathbf{-362.95}$
& $-361.22$
& $-359.09$
& $-356.36$
& $-353.68$ \\
\bottomrule
\end{tabular}

\vspace{6pt}

\begin{tabular}{@{}lrrrrr@{}}
\toprule
Application & 6 & 7 & 8 & 9 & 10 \\
\midrule
Seoul
& $-350.74$
& $-347.95$
& $-345.04$
& $-341.92$
& $-339.07$ \\
\bottomrule
\end{tabular}

\vspace{2pt}
\begin{minipage}{0.95\linewidth}
\footnotesize
\raggedright
\sloppy
\textit{Notes}. Entries are Bayesian information criterion (BIC) values~\citep{schwarz1978estimating} for VAR($p$) models with $p=1,\ldots,10$. Lower values indicate the preferred lag order, and the bold value denotes the minimum. All candidate models are estimated without an intercept or other deterministic terms, consistent with the main VAR specification. The BIC-selected $p=1$ specification is used in the baseline analysis. BIC also selects $p=1$ when an intercept is included.
\end{minipage}
\end{table}

Detailed residual diagnostics for the reported VAR specifications are provided in Supplementary Section~S4.

\subsubsection*{S1.1. Irregular-interval sensitivity}

The baseline Seoul panel contains three returns spanning 14 calendar days because three weekly index levels are unavailable. The sensitivity specification removes each long-interval return and prevents the corresponding VAR lag from crossing the resulting gap.

\begin{table}[!htbp]
\centering
\caption{Irregular Seoul intervals and their treatment in the sensitivity analysis}
\label{tab:S1_gap_dates}
\small
\setlength{\tabcolsep}{4pt}
\renewcommand{\arraystretch}{1.08}

\begin{tabular}{@{}lccc@{}}
\toprule
\shortstack{Missing weekly\\index date}
& \shortstack{Long-return\\end date}
& \shortstack{Interval\\(days)}
& \shortstack{Following VAR\\response excluded} \\
\midrule
October 2, 2017  & October 9, 2017  & 14 & October 16, 2017 \\
January 27, 2025 & February 3, 2025 & 14 & February 10, 2025 \\
October 6, 2025  & October 13, 2025 & 14 & October 20, 2025 \\
\bottomrule
\end{tabular}

\vspace{2pt}
\begin{minipage}{0.95\linewidth}
\footnotesize
\raggedright
\sloppy
\textit{Notes}. The baseline retains the return ending after each missing weekly index date. The sensitivity specification excludes each 14-day return and the following VAR response--lag pair, preventing a removed long-interval return from entering either side of the VAR(1) regression.
\end{minipage}
\end{table}

The gap-adjusted sample contains 740 returns and 736 usable VAR(1) response--lag pairs, with a companion radius of 0.975.

\begin{table}[!htbp]
\centering
\caption{Sensitivity to the three irregular 14-day return intervals}
\label{tab:S1_gap_sensitivity}
\small
\setlength{\tabcolsep}{4pt}
\renewcommand{\arraystretch}{1.08}

\begin{tabular}{@{}lrrrrr@{}}
\toprule
Specification
& \shortstack{TCI\\(\%)}
& Reversals
& $\rho_{\mathrm{NET}}$
& $\rho_{\mathrm{PR}}$
& $\rho_{\mathrm{VC}}$ \\
\midrule
Gap-adjusted sample
& 80.96
& 28
& 0.889
& 0.901
& 0.846 \\
\bottomrule
\end{tabular}

\vspace{2pt}
\begin{minipage}{0.95\linewidth}
\footnotesize
\raggedright
\sloppy
\textit{Notes}. Rank correlations are Spearman correlations with the baseline $H=10$ ranking. Direction reversals are counted relative to the baseline pairwise net-connectedness network across the 300 district pairs.
\end{minipage}
\end{table}

The irregular-interval adjustment affects finer edge directions and district ordering but leaves the aggregate connectedness level close to the baseline value. This specification is treated as a data-construction sensitivity check rather than as an alternative baseline sample.

\subsection*{S2. Rolling-window sensitivity}

\subsubsection*{S2.1. Baseline rolling series and window-length sensitivity}

The rolling analysis re-estimates the no-intercept VAR(1)--GFEVD system at $H=10$ and uses the baseline $W=52$ series to trace broad changes in system-wide housing-market connectedness at approximately annual resolution. Because the Seoul system contains 25 variables, this specification leaves only 51 effective observations per equation after lag construction. The rolling TCI is therefore interpreted as a descriptive system-level temporal profile, and its sensitivity to temporal resolution is examined using $W\in\{52,104,156,208,286,429\}$.

\begin{figure}[!htbp]
\centering
\includegraphics[width=0.90\textwidth]{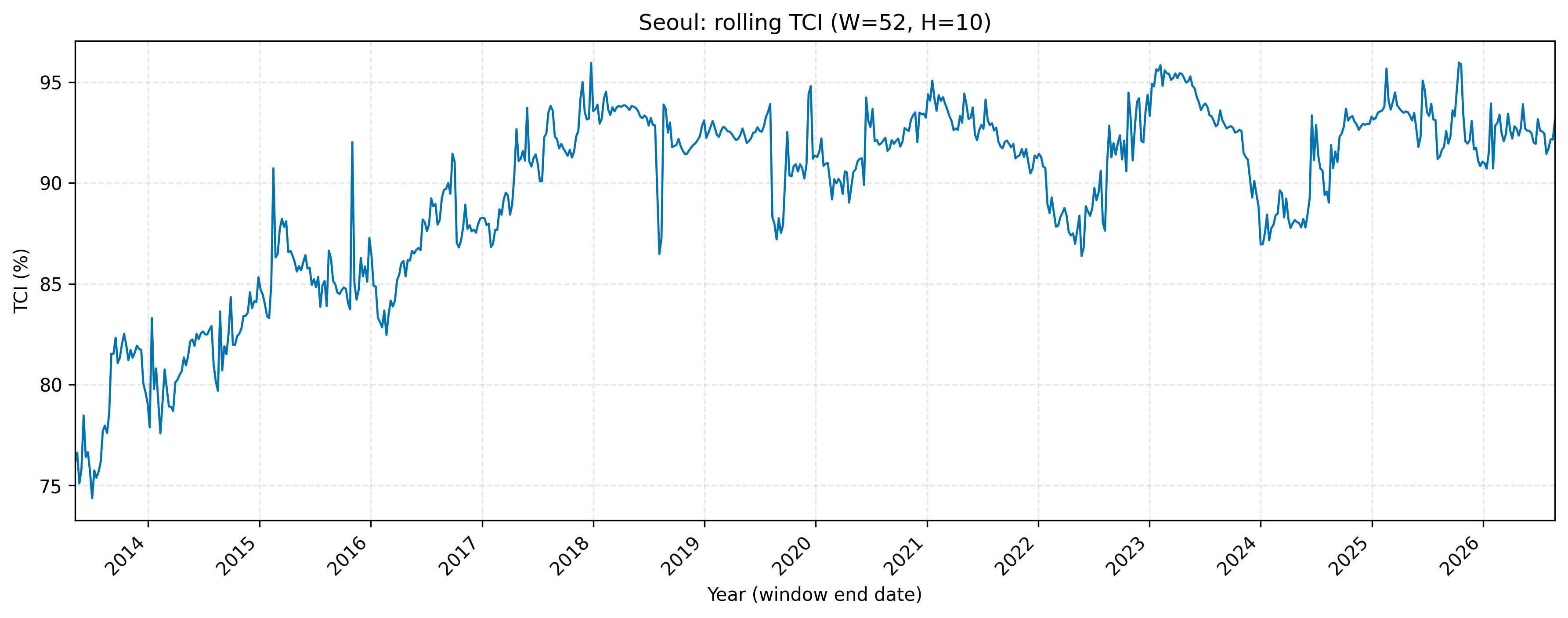}
\caption{Baseline 52-week rolling Total Connectedness Index for Seoul at $H=10$. Each estimate is plotted at the end date of its estimation window. The series traces broad changes in system-wide connectedness, while its absolute level is interpreted in light of the window-length sensitivity reported below.}
\label{fig:S2_baseline_rolling}
\end{figure}

\begin{table}[!htbp]
\centering
\caption{Sensitivity of the mean rolling Total Connectedness Index to the estimation-window length at forecast horizon $H=10$}
\label{tab:S2_window_ladder}
\scriptsize
\setlength{\tabcolsep}{3.5pt}
\renewcommand{\arraystretch}{1.08}

\begin{tabular}{@{}lrrrrrrr@{}}
\toprule
& \multicolumn{7}{c}{Estimation window $W$ (weeks)} \\
\cmidrule(lr){2-8}
Application & 52 & 104 & 156 & 208 & 286 & 429 & \shortstack{Full\\sample} \\
\midrule
Seoul (\%)
& 89.6 & 86.3 & 85.7 & 86.1 & 87.0 & 87.1 & 82.0 \\
\bottomrule
\end{tabular}

\vspace{2pt}
\begin{minipage}{0.95\linewidth}
\footnotesize
\raggedright
\sloppy
\textit{Notes}. Each entry is the mean rolling TCI obtained using the indicated window length over the full set of rolling windows available for that specification. Rolling windows advance one week at a time, and the forecast horizon is fixed at $H=10$. The full-sample column reports the corresponding full-sample TCI for the 743 Seoul observations. The full-sample TCI is 81.98\%.
\end{minipage}
\end{table}

The mean rolling TCI varies with the estimation window even though the underlying data, VAR lag, and forecast horizon are held fixed. This dependence on temporal resolution is why the rolling series is used descriptively rather than as an invariant estimate of system-wide connectedness.

\subsubsection*{S2.2. Cross-window stability of temporal variation}

\begin{figure}[!htbp]
\centering
\includegraphics[width=\textwidth]{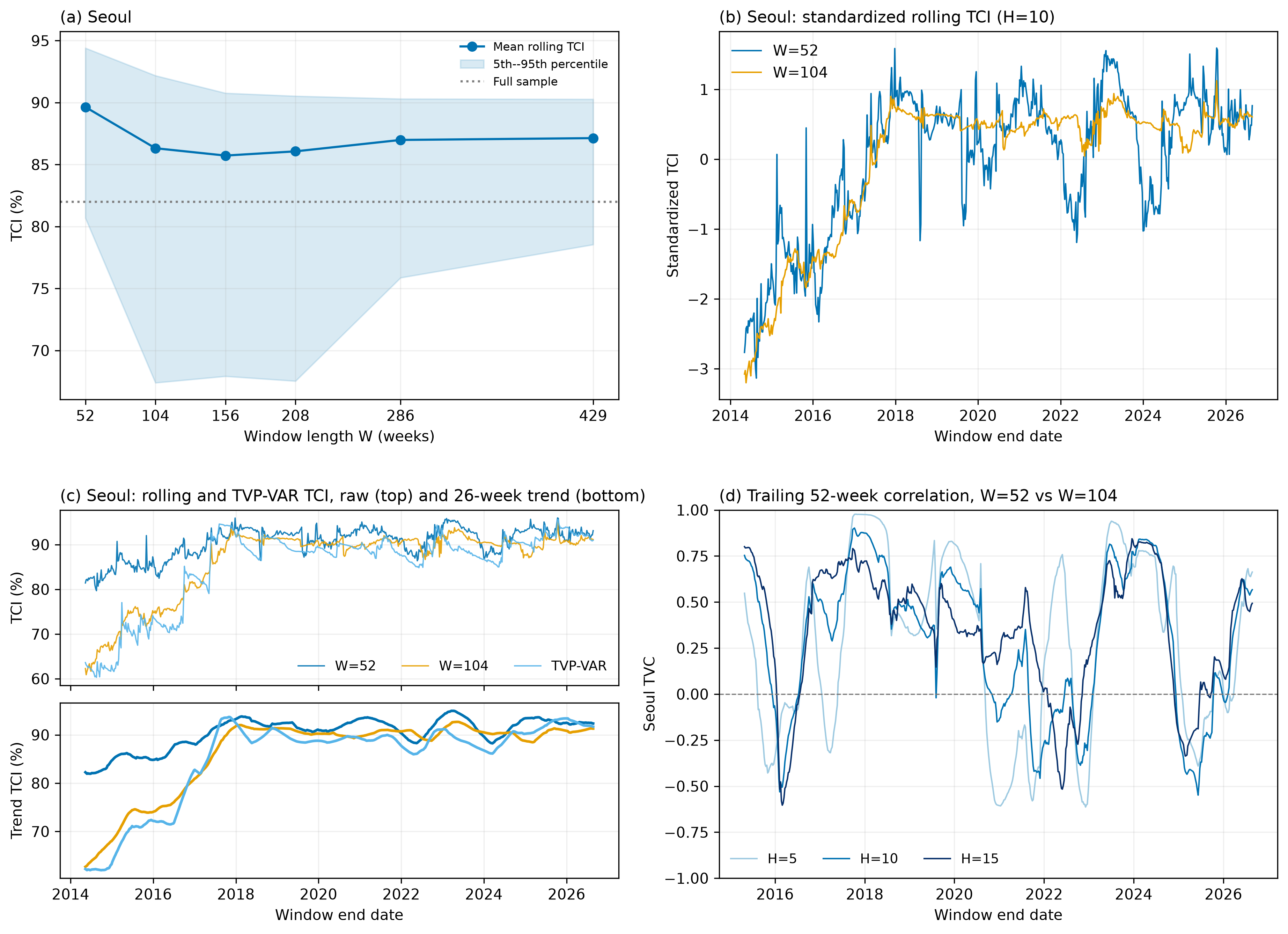}
\caption{Rolling TCI diagnostics for window-length sensitivity and temporal stability in Seoul. Panel (a) reports mean rolling TCI by window length at $H=10$, with 5th--95th percentile bands and the full-sample TCI. Panel (b) compares standardized $W=52$ and $W=104$ series. Panel (c) compares the Seoul $W=52$, $W=104$, and TVP--VAR paths as raw series and 26-week trends. Panel (d) reports trailing 52-week correlations between the $W=52$ and $W=104$ series at $H=5$, $H=10$, and $H=15$.}
\label{fig:S2_rolling_diagnostics}
\end{figure}

To distinguish sensitivity in the rolling TCI level from similarity in its temporal variation, the $W=52$ and $W=104$ series are aligned by common window-end date. Figure~\ref{fig:S2_rolling_diagnostics} compares their standardized paths, trend components, a recursive TVP--VAR benchmark, and trailing cross-window correlations. The recursive benchmark follows the forgetting-factor TVP--VAR updating logic described by \citet{antonakakis2020refined} and used in the Seoul housing application of \citet{hwang2021analyzing}. It provides an additional temporal benchmark for assessing whether the broad pattern of increases and decreases is specific to the fixed-window estimator, rather than serving as an alternative baseline estimator.

\begin{table}[!htbp]
\centering
\caption{Cross-window association between the $W=52$ and $W=104$ rolling Total Connectedness Index series}
\label{tab:S2_cross_window}
\scriptsize
\setlength{\tabcolsep}{3pt}
\renewcommand{\arraystretch}{1.08}

\begin{tabular}{@{}llrrrrrr@{}}
\toprule
Application
& $H$
& \shortstack{Align.\\$n$}
& \shortstack{Pearson\\$r$}
& \shortstack{Spearman\\$\rho_S$}
& \shortstack{Mean\\TVC}
& \shortstack{Min.\\TVC}
& $\Pr(\mathrm{TVC}>0)$ \\
\midrule
Seoul & 5  & 640 & 0.809 & 0.749 & 0.267 & $-0.613$ & 65.0\% \\
Seoul & 10 & 640 & 0.846 & 0.676 & 0.293 & $-0.549$ & 71.5\% \\
Seoul & 15 & 640 & 0.853 & 0.654 & 0.345 & $-0.603$ & 83.0\% \\
\bottomrule
\end{tabular}

\vspace{2pt}
\begin{minipage}{0.95\linewidth}
\footnotesize
\raggedright
\sloppy
\textit{Notes}. Pearson $r$ and Spearman $\rho_S$ are calculated between the $W=52$ and $W=104$ rolling TCI series after alignment by common window-end date. TVC denotes the trailing 52-week Pearson correlation between the two aligned series, and $\Pr(\mathrm{TVC}>0)$ is the share of available trailing correlations that are positive. The first 51 aligned observations have no TVC estimate by construction. At $H=10$, trailing windows of 26 and 104 weeks yield mean TVC values of 0.248 and 0.359, respectively, compared with 0.293 for the 52-week window.
\end{minipage}
\end{table}

At $H=10$, the aligned $W=52$ and $W=104$ series have Pearson and Spearman correlations of 0.846 and 0.676, respectively. Their association weakens as the comparison window becomes substantially longer, consistent with increasing temporal averaging. The 26-week trend correlations with the recursive TVP--VAR benchmark are 0.930 for $W=52$ and 0.974 for $W=104$. The comparisons indicate that the broad temporal pattern is not specific to the 52-week specification, although the absolute level and smoothness of the rolling TCI depend on temporal resolution. These results are used to characterize system-level variation; district-level temporal positions are assessed separately in Stage~4.

\subsection*{S3. Bootstrap uncertainty and edge-direction diagnostics}

This section reports full-sample residual-bootstrap rank distributions, sensitivity to the bootstrap block length, and edge-direction stability. These analyses complement the node-level results in the main text. They are distinct from Stage~4, which separately evaluates observed split-sample persistence and within-subsample bootstrap reproducibility using independently estimated early and late subsample models.

\subsubsection*{S3.1. Full-sample rank distributions}

Rank uncertainty is evaluated by jointly resampling fitted VAR residual vectors in moving blocks, recursively generating return paths from the fitted no-intercept VAR(1), and re-estimating the VAR--GFEVD network and the three node-level measures in each replication. The rank-distribution analysis uses 2,000 replications and reports the median rank, the 2.5th--97.5th percentile interval, the full-sample rank, and the frequency of top-three membership.

\begin{figure}[!htbp]
\centering
\includegraphics[width=\textwidth]{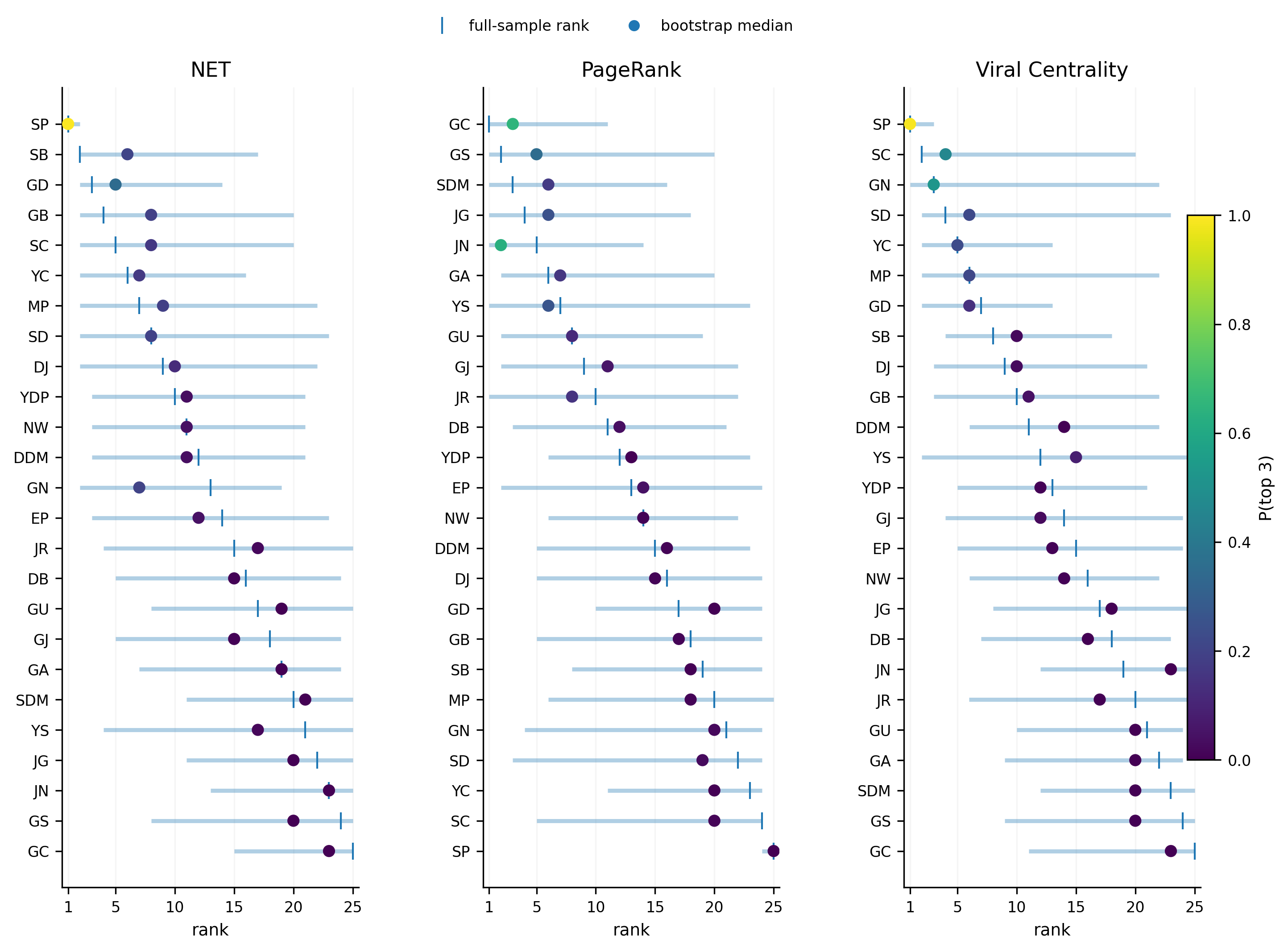}
\caption{Residual-bootstrap distributions of Seoul district ranks under NET, PageRank, and Viral Centrality. Results are based on 2,000 residual moving-block bootstrap replications with a 12-week block length. Circles indicate median bootstrap ranks, horizontal intervals show the 2.5th--97.5th percentile range, and vertical markers indicate the corresponding full-sample ranks. The shading of each marker represents the frequency of top-three membership.}
\label{fig:S3_bootstrap_ranks}
\end{figure}

The rank distributions show that reproducibility differs across districts and measures, and that exact ordinal positions should not generally be treated as fixed. The distributions are therefore used to describe rank uncertainty rather than to assign statistical significance to individual ranks.

\subsubsection*{S3.2. Full-sample block-length sensitivity}

The full-sample rank results are additionally evaluated using residual-block lengths of 6, 12, and 24 weeks. Songpa's median NET rank is 1 under all three choices, with top-three frequencies of 99.2\%, 99.3\%, and 99.1\%, respectively. Its median Viral Centrality rank is also 1 throughout, with top-three frequencies of 99.2\%, 99.1\%, and 99.0\%; at the 12-week block length, the first-rank frequency is 90.8\%. Geumcheon's median PageRank ranks are 2, 3, and 3, with corresponding top-three frequencies of 67.9\%, 65.0\%, and 52.7\%. Gangnam's median Viral Centrality ranks are 5, 3, and 3 across the three block lengths. These full-sample results show that some broad relative positions recur under residual resampling even when exact ranks remain uncertain. They do not replace the separate split-sample assessment conducted in Stage~4.

\subsubsection*{S3.3. Edge-direction stability and the 90\% backbone}

Edge-direction stability is evaluated using 10,000 residual-bootstrap replications with a three-week block selected by the automatic block-length procedure of \citet{politis2004automatic}, with the subsequent correction of \citet{patton2009correction} taken into account. The median direction-reproduction share is 0.848. Of the 300 district pairs, 80 reproduce the estimated direction in at least 95\% of replications and 122 do so in at least 90\%. The six links emphasized in Figure~1 are reproduced in 99.6--100.0\% of replications, while Songpa's 24 outgoing directions are reproduced in 91.2--100.0\%.

Under the selected three-week residual block, Songpa's NET is positive in 100.0\% of replications and Geumcheon's is negative in 99.6\%, whereas Gangnam's NET is positive in 45.3\%. These are descriptive resampling frequencies rather than posterior probabilities or multiplicity-adjusted significance tests.

For the descriptive 90\% backbone, the 122 district pairs whose estimated direction is reproduced in at least 90\% of the residual-bootstrap replications are retained. The corresponding baseline full-sample edge weights are preserved, while all other pairwise net links are set to zero. NET, PageRank, and Viral Centrality are then recomputed on this restricted network using the same definitions as in the baseline analysis.

Relative to the complete-network rankings, the backbone rankings have Spearman correlations of 0.970 for NET, 0.934 for PageRank, and 0.840 for Viral Centrality. Songpa remains first in NET and Viral Centrality and last in PageRank, while Geumcheon remains last in NET, first in PageRank, and a zero-Viral-Centrality sink. The backbone therefore preserves the broad contrasts emphasized in the main analysis while confirming that finer node ordering is more sensitive to edge selection. The 90\% backbone is a descriptive stability filter rather than a network selected by formal significance testing.

\subsection*{S4. Dynamic and specification sensitivity}

This section reports additional diagnostics for the dynamic specification and the sensitivity analyses summarized in the main text. The purpose is to document how the estimated network changes under alternative specifications rather than to select a uniquely correct dynamic model.

\subsubsection*{S4.1. Residual and common-component subspace diagnostics}

VAR(1) remains the BIC-selected reference specification, while VAR(2), VAR(4), and ridge VAR(4) are used to evaluate sensitivity to dynamic specification and regularization.

\begin{table}[!htbp]
\centering
\caption{Residual diagnostics for Seoul VAR specifications}
\label{tab:S4_lag_residual_diagnostics}
\scriptsize
\setlength{\tabcolsep}{2.2pt}
\renewcommand{\arraystretch}{1.08}

\begin{tabular}{@{}lccccccc@{}}
\toprule
Specification
& \shortstack{LB8\\rej.}
& \shortstack{LB12\\rej.}
& \shortstack{Hosking\\$Q(12)$ (df)}
& \shortstack{Hosking\\$p$}
& \shortstack{Companion\\radius}
& $\kappa(\widehat{\Sigma}_{u})$
& \shortstack{Residual df\\per equation} \\
\midrule
VAR(1)
& 22/25 & 22/25 & 17074.699 (6875) & $<0.0001$
& 0.973 & 30.365 & 717 \\
VAR(2)
& 12/25 & 16/25 & 12885.389 (6250) & $<0.0001$
& 0.980 & 34.608 & 691 \\
VAR(4)
& 3/25 & 6/25 & 8813.408 (5000) & $<0.0001$
& 0.963 & 34.959 & 639 \\
Ridge VAR(4)
& 23/25 & 23/25 & 11341.328 (5000) & $<0.0001$
& 0.969 & 29.820 & -- \\
\bottomrule
\end{tabular}

\vspace{2pt}
\begin{minipage}{0.95\linewidth}
\footnotesize
\raggedright
\sloppy
\textit{Notes}. LB8 and LB12 report the number of equations rejecting the null hypothesis of no residual autocorrelation at the 5\% level using equation-wise Ljung--Box tests~\citep{ljung1978measure} through lags 8 and 12. The Ljung--Box tests use no fitted-parameter degrees-of-freedom correction, matching the baseline residual diagnostic. The Hosking test~\citep{hosking1980multivariate} is a multivariate portmanteau test through lag 12. The residual-covariance condition number is denoted by $\kappa(\widehat{\Sigma}_{u})$. All specifications are estimated without an intercept. For ridge VAR(4), five-fold chronological blocked cross-validation selected $\lambda=78.495$. Residual degrees of freedom are not reported because the estimator is penalized.
\end{minipage}
\end{table}

Higher lag orders reduce the number of equation-level Ljung--Box rejections but do not eliminate joint residual dependence. From VAR(1) to VAR(4), the mean across-equation mean absolute residual ACF through lag 12 declines from 0.0524 to 0.0322, while the largest equation-level absolute residual ACF declines from 0.2826 to 0.1466. The multivariate Hosking test nevertheless rejects residual whiteness under all four reported specifications, while all four remain dynamically stable. These diagnostics motivate treating the alternative VAR specifications as sensitivity comparisons rather than as progressively validated approximations to a uniquely identified data-generating process.

\paragraph{Common-component subspace diagnostics.}
The centered raw panel has rank 25, whereas the centered cross-sectional-mean, PC1, and dynamic-PC1 adjusted panels have rank 24; their uncentered panels and no-intercept VAR design matrices remain rank 25 because the mean-preserving affine component is retained. The three adjusted residual covariance matrices also have rank 24, and the corresponding spectral radii restricted to the 24-dimensional stochastic subspaces are 0.914, 0.919, and 0.930, respectively. Over the common 741-observation period, the raw TCI is 82.10\%, the adjusted TCIs range from 50.60\% to 52.45\%, 91--96 of the 300 pairwise net directions are reversed, and the highlighted Songpa and Gangdong positions are unchanged.

\subsubsection*{S4.2. Additional horizon and parameter sensitivity}

Table~\ref{tab:S4_additional_sensitivity} collects sensitivity results that are summarized but not tabulated in the main text. The primary forecast-horizon comparison remains $H\in\{5,10,15\}$; $H=52$ and $H=260$ are included only as numerical long-horizon benchmarks.

\begin{table}[!htbp]
\centering
\caption{Additional sensitivity of Seoul connectedness and network rankings}
\label{tab:S4_additional_sensitivity}
\scriptsize
\setlength{\tabcolsep}{3pt}
\renewcommand{\arraystretch}{1.08}

\begin{tabular}{@{}lrrrrr@{}}
\toprule
Comparison
& \shortstack{TCI\\(\%)}
& $\rho_{\mathrm{NET}}$
& $\rho_{\mathrm{PR}}$
& $\rho_{\mathrm{VC}}$
& Reversals \\
\midrule
$H=5$ vs.\ $H=10$
& 74.99 & 0.985 & 0.944 & 0.935 & 12 \\
$H=15$ vs.\ $H=10$
& 84.61 & 0.998 & 0.986 & 0.989 & 7 \\
$H=52$ vs.\ $H=10$
& 88.04 & 0.995 & 0.962 & 0.935 & 16 \\
$H=260$ vs.\ $H=10$
& 88.35 & 0.995 & 0.960 & 0.926 & 17 \\
$\alpha=0.75$ vs.\ $\alpha=0.85$
& -- & -- & 1.000 & -- & -- \\
$\alpha=0.95$ vs.\ $\alpha=0.85$
& -- & -- & 0.998 & -- & -- \\
Intercept vs.\ no intercept
& 81.65 & 0.994 & 0.944 & 0.992 & -- \\
Unnormalized vs.\ normalized GFEVD
& -- & 0.969 & 0.975 & 0.931 & 10 \\
$H=1$ vs.\ $H=10$
& -- & 0.674 & 0.654 & 0.625 & 76 \\
\bottomrule
\end{tabular}

\vspace{2pt}
\begin{minipage}{0.95\linewidth}
\footnotesize
\raggedright
\sloppy
\textit{Notes}. Rank correlations are Spearman correlations with the corresponding baseline ranking. Horizon reversals are counted relative to the $H=10$ pairwise net-connectedness directions. Changes in $\alpha$ affect PageRank only and leave the underlying pairwise network unchanged. A dash indicates that the quantity is not applicable or is not required for the reported comparison. Irregular-interval sensitivity is reported separately in Table~\ref{tab:S1_gap_sensitivity}.
\end{minipage}
\end{table}

The horizon comparisons show that similar node rankings can coexist with changes in aggregate connectedness and pairwise direction. PageRank is nearly unchanged across the examined damping values, while the intercept and GFEVD-normalization checks leave the broad ordering comparatively similar. The $H=1$ benchmark differs more strongly and is examined separately below because it has a distinct contemporaneous structure.

\subsubsection*{S4.3. Contemporaneous topology benchmark}

At $H=1$, $\mathbf{A}_0=\mathbf{I}_N$ in Eq.~(5), so

\[
\theta^g_{ij}(1)
=
\frac{\sigma_{ij}^2}{\sigma_{ii}\sigma_{jj}}
=
\rho_{ij}^2
\]

is symmetric before row normalization. Let $S_i=\sum_k\rho_{ik}^2$. Equation~(12) then implies

\[
N_{ij}(1)
=
\frac{\rho_{ij}^2(S_i-S_j)}{S_iS_j}.
\]

Every nonzero edge is therefore oriented according to the ordering of $S_i$, so the nonzero support is acyclic. If every pair has a nonzero net difference, this support is a transitive tournament. In Seoul, the $H=1$ network has no cyclic triads, whereas the $H=10$ network contains 97 and differs in direction for 76 of the 300 district pairs. The baseline topology therefore contains lag-dependent directed structure that is absent from the contemporaneous benchmark.

\subsection*{S5. Structural benchmark, controlled tournament simulation, and Viral Centrality implementation}

\subsubsection*{S5.1. Proofs for the edge-flow decomposition and monotone-allocation benchmark}

We first derive the specialization of the least-squares edge-flow decomposition used in the main article. Let
\[
b_i=\sum_{j=1}^{N}N_{ij}
\]
and consider
\[
L(\mathbf{x})
=
\sum_{i<j}
\left[
N_{ij}-(x_i-x_j)
\right]^2
\]
subject to $\sum_i x_i=0$. The first-order condition for node $i$ is
\[
N x_i-\sum_{j=1}^{N}x_j-\sum_{j=1}^{N}N_{ij}=0.
\]
Under the zero-sum constraint, the unique solution is therefore
\[
x_i
=
\frac{b_i}{N}
=
\frac{NET_i}{100N}
=
p_i.
\]
Uniqueness follows because the complete-graph Laplacian is positive definite on the zero-sum subspace.

Define
\[
G_{ij}=p_i-p_j,
\qquad
C_{ij}=N_{ij}-G_{ij}.
\]
Since $\sum_i p_i=0$,
\[
\sum_{j=1}^{N}G_{ij}
=
Np_i
=
\sum_{j=1}^{N}N_{ij},
\]
and hence
\[
\sum_{j=1}^{N}C_{ij}=0.
\]
Moreover,
\[
\langle\mathbf{G},\mathbf{C}\rangle_F
=
\sum_{i,j}(p_i-p_j)C_{ij}
=
2\sum_i p_i\sum_j C_{ij}
=
0.
\]
Thus, $\mathbf{G}$ and $\mathbf{C}$ are orthogonal, and $\mathbf{C}$ is a divergence-free, NET-orthogonal residual that leaves every node's NET unchanged.

We next prove the monotone-allocation proposition. Index the nodes of a complete transitive tournament so that $j\rightarrow k$ whenever $j<k$, making node $N$ the unique sink. For node $i$, the PageRank equation is
\[
\pi_i
=
\frac{1-\alpha}{N}
+
\alpha
\left[
\frac{\pi_N}{N}
+
\sum_{j<i}\pi_j Q_{ji}
\right].
\]
For $i=1,\ldots,N-1$, subtraction of the equations for adjacent nodes gives
\[
\pi_{i+1}-\pi_i
=
\alpha
\left[
\pi_i Q_{i,i+1}
+
\sum_{j<i}
\pi_j
\left(
Q_{j,i+1}-Q_{j,i}
\right)
\right].
\]
Strictly positive tournament edge weights imply $Q_{i,i+1}>0$, while uniform teleportation implies $\pi_i>0$. The monotone-allocation condition makes every term in the summation nonnegative. Therefore,
\[
\pi_{i+1}>\pi_i.
\]
Because
\[
d_i^{\mathrm{in}}=i-1,
\]
PageRank and indegree induce the same ranking.

Finally, suppose that $\mathbf{C}=\mathbf{0}$ and all NET values are distinct. Then
\[
N_{ij}=p_i-p_j.
\]
Relabel the nodes so that
\[
p_1>p_2>\cdots>p_N.
\]
The resulting tournament is transitive. For $j<k<N$,
\[
Q_{j,k+1}
=
\frac{p_j-p_{k+1}}
{\sum_{\ell>j}(p_j-p_\ell)}
>
\frac{p_j-p_k}
{\sum_{\ell>j}(p_j-p_\ell)}
=
Q_{j,k}.
\]
The monotone-allocation proposition therefore applies. Since
\[
NET_i=100Np_i,
\]
PageRank and indegree induce the exact reverse of the NET ordering.

For a complete unit-weight tournament with odd $N\geq3$, let $d_i^{\mathrm{out}}$ denote the outdegree of node $i$. The row sum of the skew-symmetric pairwise-net matrix is
\[
b_i
=
\sum_j N_{ij}
=
2d_i^{\mathrm{out}}-(N-1),
\]
so that $p_i=b_i/N$. Because $\sum_i p_i=0$,
\[
\|\mathbf{G}\|_F^2
=
\sum_{i,j}(p_i-p_j)^2
=
2N\sum_i p_i^2
=
\frac{2}{N}\sum_i b_i^2.
\]
For unit edge weights,
\[
\|\mathbf{N}\|_F^2
=
N(N-1),
\]
and therefore
\[
\psi_G
=
\frac{2\sum_i b_i^2}
{N^2(N-1)}.
\]

For a tournament, the number of cyclic triads satisfies
\[
c_3
=
\binom{N}{3}
-
\sum_i
\binom{d_i^{\mathrm{out}}}{2},
\]
which gives
\[
\sum_i b_i^2
=
\frac{N^3-N-24c_3}{3}.
\]
Using
\[
\eta_{\mathrm{KBS}}
=
1-\frac{24c_3}{N^3-N},
\]
we obtain
\[
\psi_G
=
\eta_{\mathrm{KBS}}
\frac{2(N+1)}{3N}.
\]
For a transitive tournament, $c_3=0$ and $\eta_{\mathrm{KBS}}=1$, so
\[
\psi_G
=
\frac{2(N+1)}{3N}.
\]
At $N=25$, this transitive uniform-weight value is $0.6933$.

The converse does not hold. For $N>2$, a transitive uniform-weight tournament generally has a nonzero NET-orthogonal residual because its constant edge weights are not equal to the differences of the least-squares node potentials. Nevertheless, uniform outgoing allocation satisfies the monotone-allocation condition and preserves exact PageRank--indegree correspondence.

\subsubsection*{S5.2. Controlled weighted-tournament simulation}

To examine structural departures from the transitive uniform-weight benchmark, we simulate complete weighted tournaments with $N=25$. Each network starts from a transitive orientation, after which each unordered pair is independently reversed with probability 0, 0.01, 0.02, 0.05, 0.10, 0.15, 0.20, 0.30, 0.40, or 0.50. Positive edge weights are drawn independently of the orientation from mean-one lognormal distributions with target coefficients of variation of 0, 0.25, 0.50, 1, or 2. For each of the 50 parameter combinations, 1,000 replications are generated, yielding 50,000 weighted tournaments. PageRank and Viral Centrality are computed using the same specifications as in the empirical analysis. The simulation uses seed 20260930.

At the transitive uniform-weight boundary, both the PageRank--indegree and Viral Centrality--outdegree rank correlations equal one. Away from this boundary, the relationship between tournament consistency and degree--centrality alignment is not monotonic over the full range of simulated structures. Greater edge-weight heterogeneity generally weakens alignment, while substantial correspondence can still arise in tournaments with lower consistency. Tournament consistency therefore provides a useful structural descriptor but does not by itself determine the resulting centrality ranking.

The baseline Seoul network has $\eta_{\mathrm{KBS}}=0.851$ and a coefficient of variation of 0.878 for its positive pairwise-net edge weights. Among the 245 simulated networks lying within 0.025 of Seoul's tournament consistency and within 0.10 of its realized edge-weight coefficient of variation, the PageRank--indegree correlation has a median of approximately 0.833, a descriptive 2.5th--97.5th percentile range of approximately 0.654--0.952, and a maximum of approximately 0.967, compared with 0.987 in Seoul. The corresponding Viral Centrality--outdegree correlation has a median of approximately 0.863 and a descriptive 2.5th--97.5th percentile range of approximately 0.654--0.959, with the Seoul value of 0.935 lying within this range.

These comparisons are controlled structural benchmarks rather than stochastic null tests for the Seoul district positions. In particular, simulated edge weights are generated independently of orientation, whereas the directions and weights in the empirical Seoul network are jointly obtained from the estimated pairwise net-connectedness matrix. The simulated frequencies and ranges are therefore not interpreted as $p$-values or significance statements about the empirical network positions.

\subsubsection*{S5.3. Numerical convergence of Viral Centrality}

Numerical convergence is stable across the tested Viral Centrality specifications. Across the evaluated combinations of application, forecast horizon, and network construction, no calculation reaches the 10,000-iteration limit. Tightening the convergence tolerance from $10^{-10}$ to $10^{-12}$ changes the Viral Centrality score by at most $9.52\times10^{-10}$, and the resulting rank correlations are 1.000 in every comparison. The reported Viral Centrality rankings are therefore insensitive to the numerical stopping tolerance within the examined range.

\subsubsection*{S5.4. Network-construction sensitivity of Viral Centrality}

For every non-sink node, row normalization makes the outgoing weights in $\mathbf{B}(H)$ sum to one. Viral Centrality therefore preserves relative outgoing allocation and downstream path structure rather than absolute outgoing net-connectedness magnitude. The spectral radius of $\mathbf{B}(H)$ declines from 1.000 at $H=5$ to 0.685 at $H=10$, 0.560 at $H=15$, 0.397 at $H=52$, and approximately 0.386 over $H=104$--$260$. Geumcheon is a sink at every evaluated horizon from $H=10$ onward. The number of cyclic triads declines from 141 at $H=5$ to 97 at $H=10$, 47 at $H=15$, 22 at $H=52$, and 17 over $H=104$--$260$. These quantities describe changes in the network operator and are not used as convergence conditions for the Viral Centrality iteration.

To assess sensitivity to edge selection and weight construction, Viral Centrality is recomputed after retaining only the 50\%, 20\%, 10\%, and 5\% highest-weight edges of $\mathbf{A}^{+}(H)$ and under the two alternative weight mappings reported in Table~\ref{tab:S5_vc_sensitivity}.

\begin{table}[!htbp]
\centering
\caption{Additional sensitivity of Viral Centrality to edge thresholding and weight mapping at $H=10$}
\label{tab:S5_vc_sensitivity}
\small
\setlength{\tabcolsep}{5pt}
\renewcommand{\arraystretch}{1.08}

\begin{tabular}{@{}lrrrr@{}}
\toprule
\multicolumn{5}{c}{Thresholded network} \\
\midrule
Application & Top 50\% & Top 20\% & Top 10\% & Top 5\% \\
\midrule
Seoul & 0.827 & 0.565 & 0.566 & 0.471 \\
\bottomrule
\end{tabular}

\vspace{6pt}

\begin{tabular}{@{}lrr@{}}
\toprule
\multicolumn{3}{c}{Weight mapping} \\
\midrule
Application & $\beta=A^+$ & $\beta=A^+/\max A^+$ \\
\midrule
Seoul & 0.819 & 0.878 \\
\bottomrule
\end{tabular}

\vspace{2pt}
\begin{minipage}{0.95\linewidth}
\footnotesize
\raggedright
\sloppy
\textit{Notes}. Entries are Spearman rank correlations with the baseline Viral Centrality ranking at $H=10$. Thresholded networks are row-normalized after edge selection. The alternative weight mappings are sensitivity specifications rather than calibrated diffusion probabilities.
\end{minipage}
\end{table}

Songpa remains first under the reported threshold and weight-mapping alternatives, although the finer district ordering becomes increasingly sensitive as the network is sparsified or the edge mapping is changed. These exercises are treated as implementation and network-construction sensitivities rather than as additional candidate-level diagnostic rules.

\subsection*{S6. Low-connectedness stress test: New York City}

The New York City application is used as a low-connectedness stress test rather than as an external replication of the Seoul district-level ordering. It uses weekly transaction-based residential price-per-square-foot averages for the five boroughs over 2010--2020. Transactions with valid sale dates, positive sale prices, positive gross square feet, and valid borough identifiers are retained, with no interpolation of empty weekly cells. The resulting complete-case return panel contains 571 weekly observations. All five return series reject the unit-root null at the 1\% level under both ADF and PP tests, and BIC selects VAR(1). Because the data construction, spatial aggregation, network size, and institutional setting differ materially from the Seoul application, the connectedness and centrality magnitudes are not compared directly across the two cities.

\begin{table}[!htbp]
\centering
\caption{Low-connectedness stress-test diagnostics for New York City}
\label{tab:S6_nyc_stress_test}
\small
\setlength{\tabcolsep}{5pt}
\renewcommand{\arraystretch}{1.08}

\begin{tabular}{@{}L{0.27\linewidth}L{0.67\linewidth}@{}}
\toprule
Diagnostic & Result \\
\midrule
Network and sample size
& 5 boroughs; 571 weekly return observations \\
Full-sample TCI
& 0.192\% \\
Bootstrap rank uncertainty
& The 2.5th--97.5th percentile intervals span ranks 1--5 for every borough and node-level measure \\
Median direction-reproduction share
& 0.69 \\
Pairwise directions reproduced in at least 90\% of replications
& 0 of 10 \\
Intercept sensitivity
& TCI changes from 0.192\% to 0.196\%; rank correlations are 0.900 for NET, 0.800 for PageRank, and $-0.600$ for Viral Centrality \\
Rolling-TCI window sensitivity
& Mean TCI declines from 17.6\% at $W=52$ to 1.9\% at $W=429$ \\
Data-construction caveat
& Manhattan has a median of 10 qualifying transactions per week, with 46.0\% of weeks containing fewer than ten \\
\bottomrule
\end{tabular}
\end{table}

The full-sample New York City rankings are treated as unstable point estimates. The full-sample TCI is only 0.192\%, the bootstrap rank intervals span positions 1--5 for every borough and node-level measure, and none of the ten pairwise directions is reproduced in at least 90\% of the residual-bootstrap replications. The exercise therefore illustrates how the same computational framework behaves in a weakly connected setting, but it is not used as external validation of the Seoul district-level ordering or diagnostic profiles.

\end{document}